\documentclass[aps,preprint,showpacs,preprintnumbers,amsmath,amssymb]{revtex4}
\usepackage{amsmath,mathrsfs,amsbsy,color,graphicx,bm,amsthm,amsfonts}
\usepackage{units}
\usepackage{bbm}
\usepackage{times}
\usepackage{dcolumn}
\usepackage{mathrsfs}
\usepackage{amsmath,amssymb,epsfig}
\usepackage{amsmath}
\newcommand{\udots}{\mathinner{\mskip1mu\raise1pt\vbox{\kern7pt\hbox{.}}
\mskip2mu\raise4pt\hbox{.}\mskip2mu\raise7pt\hbox{.}\mskip1mu}}
\begin{document}

\title{Would quantum coherence be increased by curvature effect in de Sitter space?}
\author{Shu-Min Wu$^1$\footnote{smwu@lnnu.edu.cn}, Chun-Xu Wang$^1$, Dan-Dan Liu$^1$, Xiao-Li Huang$^1$\footnote{ huangxiaoli1982@foxmail.com}, Hao-Sheng Zeng$^2$\footnote{hszeng@hunnu.edu.cn (corresponding author)}}
\affiliation{$^1$ Department of Physics, Liaoning Normal University, Dalian 116029, China\\
$^2$ Department of Physics, Hunan Normal University, Changsha 410081, China
}


\begin{abstract}
We study the quantum coherence in de Sitter space for the bipartite system of Alice and Bob who initially share an entangled state between the two modes of a free massive scalar field. It is shown that the space-curvature effect can produce both local coherence and correlated coherence, leading to the increase of the total coherence of the bipartite system. These results are sharp different from the Unruh effect or Hawking effect, which, in the single mode approximation, cannot produce local coherence and at the same time destroy correlated coherence, leading to the decrease of the total coherence of the bipartite systems.
Interestingly, we find that quantum coherence has the opposite behavior
compared with the quantum correlation in de Sitter space.
We also  find that quantum coherence is most severely affected by the curvature effect of de Sitter space for the cases of conformal invariance and masslessness.
Our result reveals the difference between the curvature effect in the de Sitter space and the Unruh effect in Rindler spacetime or the Hawking effect in black hole spacetime on quantum coherence.
\end{abstract}

\vspace*{0.5cm}
\maketitle
\section{Introduction}
The coherent superposition of states is one of the characteristic features, which mark the
departure of quantum mechanics from the classical physics, if not the most essential one \cite{L1}.
Quantum entanglement and quantum coherence are arguably the two cardinal attributes of quantum theory, originating from the tensor product structure and the quantum superposition principle, respectively \cite{L2,L3,L4}. Even the phenomena of quantum  entanglement and discord can also be assigned to the nonlocal superpositions of quantum states \cite{L5,LL5,LLL5}.  Analogous to quantum entanglement and discord,
quantum coherence is also a powerful resource, which is widely used in such as quantum information, quantum metrology, thermodynamics, solid-state physics and biological systems \cite{L6,L7,L9,L10,L11,L12,L13,L14,L15,L16}.
Despite the fundamental importance of quantum coherence, it  received a lot of attention until
Baumgratz $et$ $al.$ put forward quantification of
coherence, including the $l_1$ norm coherence and the relative entropy of coherence \cite{L17}.

Quantum mechanics and relativity theory
represent two fundamentals of modern physics, but unification of the two theories remains an open problem.  Much effort have been put forward to bridge
the gap between them, which gives birth to quantum field theory (QFT).
The important predictions in QFT are the Unruh effect, the Hawking radiation, and the generation of entanglement entropy in an expanding universe and de Sitter space, which tells us that the vacuum of the field depends on the observer \cite{L18,L19,L20,L20L,L20LL}. It has been found that (i) quantum entanglement, discord, and coherence
decrease monotonically with the increase of acceleration in Rindler spacetime \cite{L21,L22,L23,L24,L25,L26,L27,L28,L29,L30,L31,L32,L33,L34}; (ii)
quantum entanglement, discord, and coherence  suffer from
a degradation due to the Hawking radiation in the background
of black holes \cite{QWE81,QWE82,L35,L36,L37,L38,L39,L40}; (iii) quantum entanglement and discord decrease monotonically with the growth of the curvature in de Sitter space \cite{LL42,L42,L43,Lqw43}. From these researches, we find that quantum coherence in both Rindler and black hole spacetimes decreases monotonically. A question naturally arises: How does curvature effect affect  quantum coherence in de Sitter space?

Our universe,  undergoing an accelerated expansion, can be approximated
by an exponential de Sitter space with a positive cosmological constant.  It is also widely believed that de Sitter space is a better description of the early stage of universe inflation. The importance comes from the fact that the de Sitter space is the unique maximally symmetric curved space.
Moreover, it is one of the cornerstones of the accelerating expansion of the cosmology that the large scale structure of our
universe and temperature fluctuations of the cosmic microwave background
(CMB) originate from quantum fluctuations during the initial inflationary
era. Therefore, the investigation  in curved
spacetime may itself lead to a more precise prediction for
cosmological observations and offer a better understanding of the initial
stages of our universe. To this aim,
quantum entanglement in  de Sitter space  has been widely studied \cite{LL42,L42,L43,Lqw43,L44,LL44,LLL44}.
As quantum coherence can  reflect nonclassical world better than quantum entanglement, the study of quantum coherence in the maximally symmetric curved space
has a deeper understanding of the quantum cosmology.

In this paper, we study the quantum coherence of an  entangled state between two free field modes
of a massive scalar field in de Sitter space.
Assume that two observers: Alice in a global chart and Bob in an open chart of de Sitter space, our aim is to study how the space background of de Sitter space influences the quantum coherence of the two-mode system observed by Alice and Bob.
Interestingly, we find that the quantum coherence increases monotonically under the curvature effect of de Sitter space. In other words, quantum coherence has the opposite behavior compared with the quantum correlation (entanglement and discord) for the two field modes in de Sitter space \cite{L42,L43}. Our result also reveals the difference of the de Sitter space with the Rindler or black hole spacetimes; in the latter two spacetimes quantum coherence decreases monotonically with the strengthening of Unruh  effect \cite{L28,L29} or Hawking effect \cite{L40}.

The paper is organized as follows. In Sec. II, we review briefly the quantization of the free massive scalar field in de Sitter space. In Sec. III, we study the behaviors of quantum coherence in de Sitter space. The last section is devoted to a brief conclusion.
\section{Quantization of scalar fields in de Sitter space \label{GSCDGE}}
\begin{figure}
\begin{minipage}[t]{0.5\linewidth}
\centering
\includegraphics[width=1.75in]{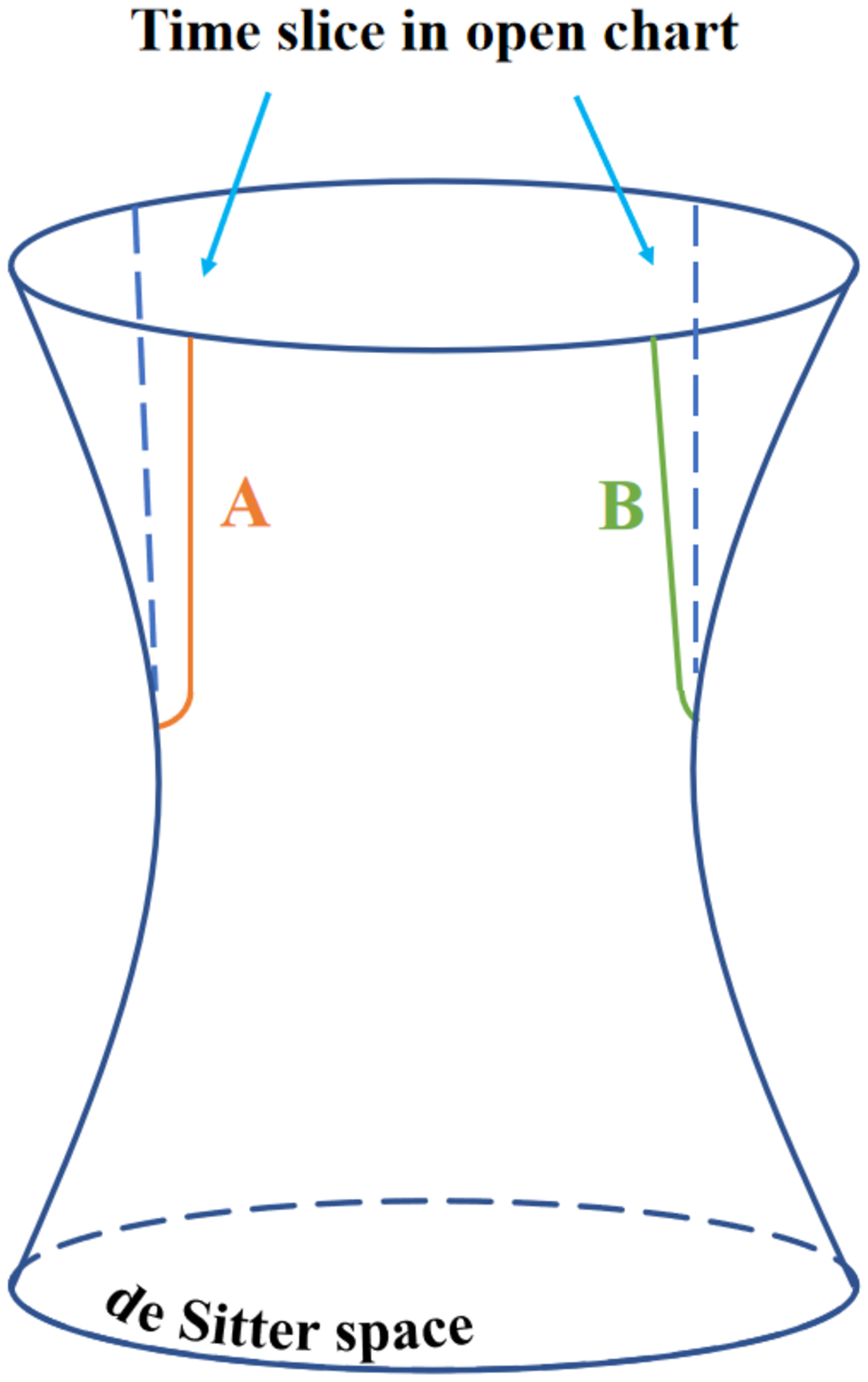}
\label{fig:side:a}
\end{minipage}%
\begin{minipage}[t]{0.5\linewidth}
\centering
\includegraphics[width=3in]{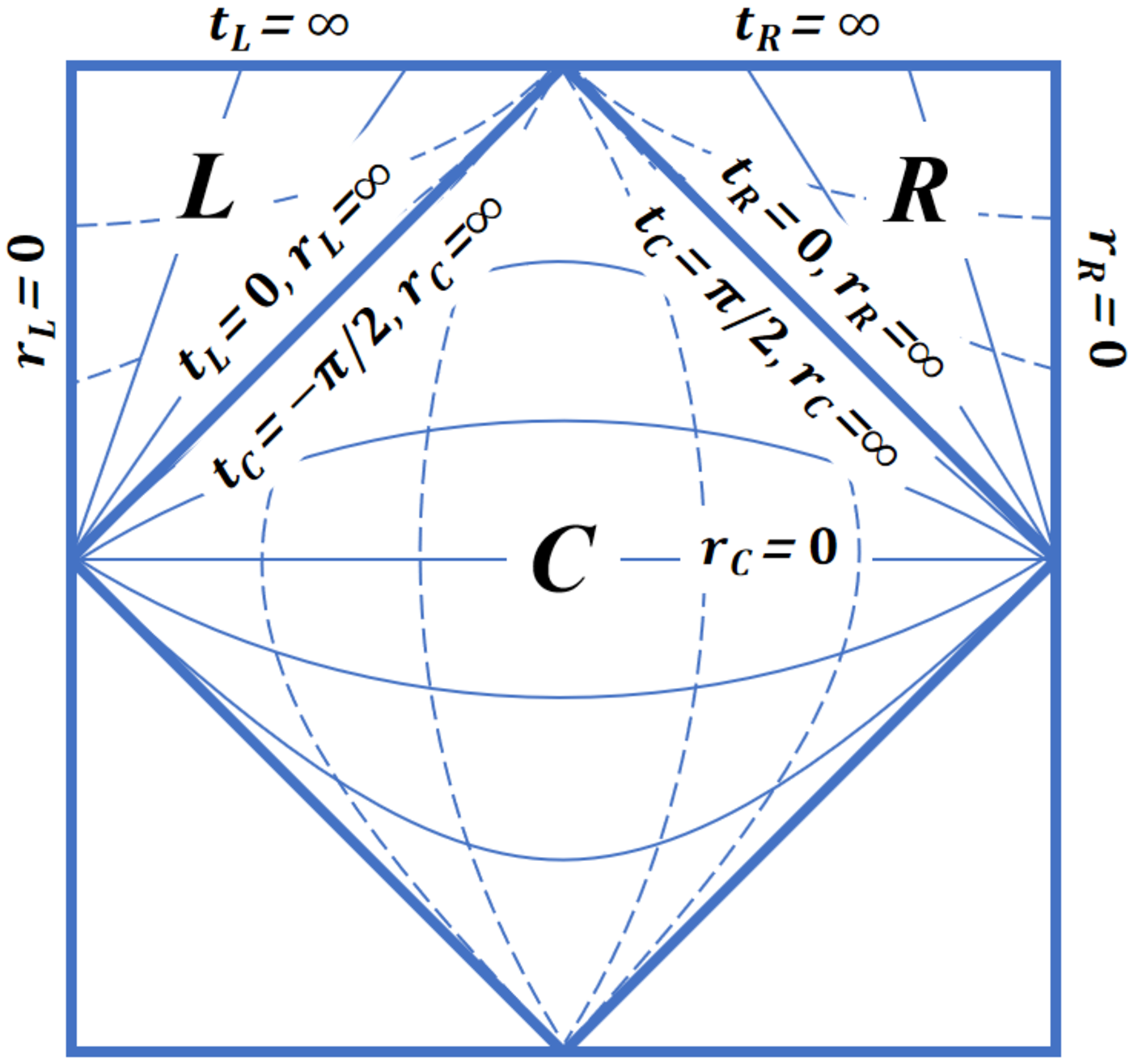}
\label{fig:side:a}
\end{minipage}%

\caption{De Sitter space and its Penrose diagram. The  $R$ and $L$ regions are the causally disconnected open charts of de Sitter space.
 }
\label{Fig1}
\end{figure}

The quantization of scalar fields in de Sitter space has been described many times before. Here we only give a brief review following the references \cite{L44,L45}. We consider a free scalar field $\phi$ with mass $m$ in de Sitter space with metric $g_{\mu\nu}$ .
The action of the scalar field is given by
\begin{eqnarray}\label{w1}
S=\int d^4 x\sqrt{-g}\left[\,-\frac{1}{2}\,g^{\mu\nu}
\partial_\mu\phi\,\partial_\nu \phi
-\frac{m^2}{2}\phi^2\,\right]\,.
\end{eqnarray}
The metric for the coordinate system that covers the whole Euclidean de Sitter space is given by
\begin{eqnarray}\label{S30e}
ds^{2}_{E}=H^{-2}[d\tau^{2}+\cos^{2}\tau(d\rho^{2}+\sin^{2}\rho d\Omega^{2})].
\end{eqnarray}
The Lorentzian region can be divided into three open charts of de Sitter space that consist of three parts: $C$, $R$ and $L$ as shown in Fig\ref{Fig1} \cite{L44,L45}, whose coordinates are related to the fundamental ones via the relationships
\begin{eqnarray}
t_{R}&=&i(\tau-\frac{\pi}{2})       \qquad \qquad(t_{R}\geq0),\notag\\
r_{R}&=&i_{\rho}     \qquad \qquad  \qquad \qquad       (r_{R}\geq0),\notag\\
t_{C}&=&\tau  \qquad\qquad\qquad    \qquad     (\frac{\pi}{2}\geq t_{C}\geq-\frac{\pi}{2}),\notag\\
r_{C}&=&i(\rho-\frac{\pi}{2}) \qquad\qquad  (\infty>r_{C}>-\infty),\notag\\
t_{L}&=&i(-\tau-\frac{\pi}{2})\qquad \qquad (t_{L}\geq0),\notag\\
r_{L}&=&i_{\rho}      \qquad  \qquad \qquad \qquad     (r_{L}\geq0),
\end{eqnarray}
and their metrics are given, respectively, by
\begin{eqnarray}\label{w2}
ds^2_C&=&H^{-2}\left[dt^2_C+\cos^2t_C\left(-dr^2_C+\cosh^2r_C\,d\Omega^2\right)
\right]\,,\nonumber\\
ds^2_R&=&H^{-2}\left[-dt^2_R+\sinh^2t_R\left(dr^2_R+\sinh^2r_R\,d\Omega^2\right)
\right]\,,\nonumber\\
ds^2_L&=&H^{-2}\left[-dt^2_L+\sinh^2t_L\left(dr^2_L+\sinh^2r_L\,d\Omega^2\right)
\right]\,,
\end{eqnarray}
where  $H^{-2}$ is the square of Hubble radius and $d\Omega^2$ is the metric on the two-sphere.
The $L$  region is causally disconnected from the  $R$ region in de Sitter space.
The $C$ region is covered by the coordinate $(t_C, r_C)$.
Since the $L$ and $R$  regions  are completely symmetric, we write their coordinates by $(t,r)=(t_R, r_R)$ or $(t_L, r_L)$, respectively.

Solving the Klein-Gordon equation in different open regions, one can obtain
\begin{eqnarray}\label{w3}
u_{\sigma p\ell m}(t_{R(L)},r_{R(L)},\Omega)&\sim&\frac{H}{\sinh t_{R(L)}}\,
\chi_{p,\sigma}(t_{R(L)})\,Y_{p\ell m} (r_{R(L)},\Omega)\,,\qquad \nonumber\\
-{\rm\bf L^2}Y_{p\ell m}&=&\left(1+p^2\right)Y_{p\ell m}\,,
\end{eqnarray}
with $Y_{p\ell m}$ being harmonic functions on the three-dimensional hyperbolic space [see appendix A].
The positive frequency mode functions $\chi_{p,\sigma}(t_{R(L)})$ that support on the $R$ and $L$  regions are defined by
\begin{eqnarray}\label{w4}
\chi_{p,\sigma}(t_{R(L)})=\left\{
\begin{array}{l}
\frac{e^{\pi p}-i\sigma e^{-i\pi\nu}}{\Gamma(\nu+ip+\frac{1}{2})}P_{\nu-\frac{1}{2}}^{ip}(\cosh t_R)
-\frac{e^{-\pi p}-i\sigma e^{-i\pi\nu}}{\Gamma(\nu-ip+\frac{1}{2})}P_{\nu-\frac{1}{2}}^{-ip}(\cosh t_R)
\,,\\
\\
\frac{\sigma e^{\pi p}-i\,e^{-i\pi\nu}}{\Gamma(\nu+ip+\frac{1}{2})}P_{\nu-\frac{1}{2}}^{ip}(\cosh t_L)
-\frac{\sigma e^{-\pi p}-i\,e^{-i\pi\nu}}{\Gamma(\nu-ip+\frac{1}{2})}P_{\nu-\frac{1}{2}}^{-ip}(\cosh t_L)
\,,
\label{solutions}
\end{array}
\right.
\end{eqnarray}
where $P^{\pm ip}_{\nu-\frac{1}{2}}$ are the associated Legendre functions,
and  $\sigma=\pm 1$  distinguishes two independent
solutions for each region. These solutions can be normalized by the factor
\begin{eqnarray}\label{w5}
N_{p}=\frac{4\sinh\pi p\,\sqrt{\cosh\pi p-\sigma\sin\pi\nu}}{\sqrt{\pi}\,|\Gamma(\nu+ip+\frac{1}{2})|}\,,
\end{eqnarray}
where the eigenvalue $p$ normalized by $H$ is a positive real parameter. As the effect of the curvature of the three-dimensional hyperbolic
space starts to appear when $p$ approaches to $1$, we can regard $p$ as the description of the curvature degree of the de Sitter space. The curvature effect gets stronger as $p$ becomes smaller than $1$ \cite{L42,L43,Lqw43}. Therefore, $p$ can be
considered as the curvature parameter of the de Sitter space.
The mass parameter $\nu$ is defined by $\nu=\sqrt{\frac{9}{4}-\frac{m^2}{H^2}}$.
We will later
comment on the parameter region:  $0\leq\frac{m^2}{H^2}\leq\frac{9}{4}$ to make the discussion clear. When $\nu=\frac{1}{2}$ (or $\frac{m^2}{H^2}=2$), we have a conformally coupled scalar field and the system is conformally invariant. The minimally
coupled massless scalar field corresponds to $\nu=\frac{3}{2}$.

One can use the eigenfunctions of Eq.(\ref{w3}) to expand the scalar field as
\begin{eqnarray}\label{w6}
\hat\phi(t,r,\Omega)
=\frac{H}{\sinh t}\int dp \sum_{\sigma,\ell,m}\left[\,a_{\sigma p\ell m}\,\chi_{p,\sigma}(t)
+a_{\sigma p\ell -m}^\dagger\,\chi^*_{p,\sigma}(t)\right]Y_{p\ell m}(r,\Omega)
\,,
\end{eqnarray}
where $a_{\sigma p\ell m}$ ($a_{\sigma p\ell m}^\dagger$) is the annihilation (creation) operator. The Bunch-Davies vacuum $|0\rangle_{\rm BD}$
can be defined as $a_{\sigma p\ell m}|0\rangle_{\rm BD}=0$. For simplicity, hereafter, we omit the indices $p$, $\ell$, $m$ of the mode functions and operators.
For example, the mode functions and the associated Legendre functions can be  rewritten in  the simple forms $\chi_{p,\sigma}(t)\rightarrow\chi^{\sigma}$,
$P_{\nu-1/2}^{ip}(\cosh t_{R,L})\rightarrow P^{R, L}$, and
$P_{\nu-1/2}^{-ip}(\cosh t_{R,L})\rightarrow P^{R*, L*}$.

Next, we consider the positive frequency mode functions
corresponding to the the R-vacuum (L-vacuum), which has supported only on the $R$ ($L$)
region.
We define the mode functions in the each $R$ and $L$ regions  of open charts of  de Sitter space by
\begin{eqnarray}
\varphi^q=\left\{
\begin{array}{ll}
\frac{|\Gamma(1+ip)|}{\sqrt{2p}}P^q~&\mbox{in region}~q\,,
\\
0~ &\mbox{in the opposite region}\,,
\end{array}
\right.
\label{varphi}
\end{eqnarray}
where $q=(R, L)$.
Since the field should be the same under the different mode functions, we can obtain
\begin{eqnarray}
\phi(t)=a_\sigma\,\chi^\sigma+a_\sigma^\dag\,\chi^\sigma{}^*
=b_q\,\varphi^q+b_q^\dag\,\varphi^q{}^*\,.
\label{fo}
\end{eqnarray}
Where we have introduced the new annihilation and creation operators ($b_q,b_q^\dag$), and the corresponding vacuum is defined as $b_R|0\rangle_R=0$ (R-vacuum) and $b_L|0\rangle_L=0$ (L-vacuum). In terms of the Bogoliubov transformations between operators ($a_q,a_q^\dag$) and ($b_q,b_q^\dag$), one can find that the Bunch-Davies vacuum state and  the single particle excitation state can be written as
\cite{L42}
\begin{eqnarray}\label{w9}
|0\rangle_{\rm BD}=\sqrt{1-|\gamma_p|^2}\,\sum_{n=0}^\infty\gamma_p^n|n\rangle_{\rm R}|n\rangle_{\rm L}\,,
\end{eqnarray}
\begin{eqnarray}\label{w10}
|1\rangle_{\rm BD}=\frac{1-|\gamma_{p}|^{2}}{\sqrt{2}}\sum_{n=0}^{\infty}\gamma^{n}_p\sqrt{n+1}[|(n+1)\rangle_{\rm L}|n\rangle_{\rm R}+|n\rangle_{\rm L}|(n+1)\rangle_{\rm R}],
\end{eqnarray}
with the parameter $\gamma_p$ given by
\begin{eqnarray}\label{kkk}
\gamma_p = i\frac{\sqrt{2}}{\sqrt{\cosh 2\pi p + \cos 2\pi \nu}
 + \sqrt{\cosh 2\pi p + \cos 2\pi \nu +2 }}\,.
\label{gammap2}
\end{eqnarray}

Now we expound the claim that $p$ can be considered as the curvature parameter of the de Sitter space. By using Eq.(\ref{w9}), the reduced density matrix from Bunch-Davies basis to the basis of the open chart $L$ region is found to be
\begin{eqnarray}\label{qws119}
\rho_{\rm L}=\rm{Tr_R(|0\rangle_{BD}\langle0|)}=(1-|\gamma_{p}|^{2})\sum_{n=0}^{\infty}|\gamma_{p}|^{2n}| n\rangle_{\rm L}\langle  n|.
\end{eqnarray}
For the conformal invariance ($\nu=1/2$) and masslessness ($\nu=3/2$), we obtain $|\gamma_p|=e^{-\pi p}$. Then the reduced density matrix is given by
\begin{eqnarray}\label{qws20}
\rho_{\rm L}=(1-e^{-2\pi p})\sum_{n=0}^{\infty}e^{-2\pi pn}| n\rangle_{\rm L}\langle  n|.
\end{eqnarray}
Similarly, we can also get the reduced density matrix of Minkowski vacuum in region I of  Rindler spacetime
\begin{eqnarray}\label{qws21}
\rho_{\rm I}=(1-e^{-2\pi \frac{\omega}{a}})\sum_{n=0}^{\infty}e^{-2\pi n\frac{\omega}{a}}| n\rangle_{\rm I}\langle  n|,
\end{eqnarray}
where $a$ is the acceleration \cite{L21,L28}. Comparing Eq.(\ref{qws20}) with Eq.(\ref{qws21}), we obtain the relation $p=\frac{\omega}{a}$, which means that the increasing in acceleration is related to the decreasing of $p$ with $a\rightarrow \infty$ corresponding to $p\rightarrow0$. For the more general $\nu$, the comparing between Eq.(\ref{qws119}) and Eq.(\ref{qws21}) gives $a=-\pi\omega/\ln|\gamma_{p}|$. For the given $\nu$ and $\omega$, one can find that $a$ is a monotonic decreasing function of $p$. From the view of this comparison, we say that $p$ is the description of the curvature of de Sitter space.
The smaller the $p$, the larger the curvature. In fact, this claim has been used already \cite{L42,L43,Lqw43}.

\section{  Quantum coherence in de Sitter space \label{GSCDGE}}
Quantum  coherence, being at the heart of interference
phenomena, originates  from the quantum state superposition principle and plays an important role in physics as it enables applications which are impossible within ray optics or classical mechanics \cite{L3}. The rise of quantum mechanics as a unified picture of particles and waves strengthened the prominent role of quantum coherence in physics. Indeed, quantum coherence underlies phenomena such as multipartite interference and is the precondition of quantum correlation. In this section, we study the effect of space curvature of de Sitter space on quantum coherence. For comparison, we quantify quantum coherence in terms of two kinds of measures: the $l_1$ norm coherence and the relative entropy of coherence \cite{L17}.
In a given reference basis $\{|i\rangle\}_{i=1,\ldots,n}$ of a $n$-dimensional system, the $l_1$-norm of quantum coherence is defined as the sum of the absolute values of all the off-diagonal elements of the system density matrix $\rho$ \cite{L17}
\begin{equation}\label{w15}
C_{l_1}(\rho)=\sum_{i\neq j}|\rho_{i,j}|,
\end{equation}
and the relative
entropy of quantum coherence is given by
\begin{equation}\label{w155}
C_{\rm RE} \left( \rho  \right) = S\left( {\rho
_{\rm{diag}} } \right) - S\left( \rho  \right),
\end{equation}
where $S(\rho)$ is the von Neumann entropy of quantum state $\rho$, and $\rho_{\rm diag}$ is the state obtained from $\rho$ by deleting all off-diagonal elements.

It should be emphasized that quantum coherence depends on the choice of reference basis. Using different reference basis to represent the same quantum state
may lead to different values of coherence. In practice, the reference basis may be dictated by the physics of the problem under consideration. For example, one may
focus on the energy eigenbasis when addressing coherence in the transport phenomena and thermodynamics. In the quantum description of Young's two-slit interference, the
path basis is favorable. In this paper, we are in the particle number representation to study the dynamics of coherence for the scalar fields in de Sitter space.
In quantum optics, the coherent superposition of number states with different number of photons is very important, which plays important roles in various optical interference settings \cite{Scully}. The well-known coherent states and squeezed states of optical fields are the typical examples of this coherent superposition. For the two-mode optical fields, the coherent superposition in the photon-number bases can give rise to entanglement between the photons of the two modes (such as the two-mode squeezed states), which is also a kind of important resource. Quantum coherence is closely related to the quantum interference. The coherence of superposition between photon-number states gives rise to usually the nonuniform distribution
of optical intensity with respect to position or time, i.e., produce interference fringes, which can be detected through suitable settings. All this mature technologies in quantum optics provide the warrant for us to explore the counterpart for the scalar fields in de Sitter space.

As for the detection of quantum coherence in a given reference basis, it may be done through the so-called tomography of quantum states. For a quantum state $\rho_{2}$ of the qubit system, it contain three independent parameters and may be expressed as $\rho_{2}=\frac{1}{2}(\hat{I}+\sum_{i=1}^{3}b_{i}\hat{\sigma}_{i})$ with $b_{i}={\rm Tr}(\rho_{2}\hat{\sigma}_{i})$. It suffices to measure the average of the three Pauli operators $\hat{\sigma}_{i}$ ($i=1,2,3 $). Generically, for the $n$ level systems, the density matrix may be written as \cite{Gunter} $\rho_{n}=\frac{1}{n}\hat{I}+\frac{1}{2}\sum_{i=1}^{n-1}(\lambda_{i}\hat{\lambda}_{i})$, where $\hat{\lambda}_{i}$ ($i=1,2,\ldots,n-1$) are the generating operators of the group $SU(n)$ and $\lambda_{i}={\rm Tr}(\rho_{n}\hat{\lambda}_{i})$. This is to say, we need only measuring the average of the $n-1$ generating operators, then we can reconstruct the quantum state $\rho_{n}$. Note that the quantum coherence of qubit systems via tomography has been measured in recent experiments \cite{Xu2020,Wang2017}.

Consider two observers, Alice and Bob, who initially share an entangled state between the two modes of a free massive scalar field in de Sitter space
\begin{eqnarray}\label{w12}
|\Phi\rangle_{\rm {AB}}=\cos\theta|0_{\rm A}\rangle_{\rm{BD}}|0_{\rm B}\rangle_{\rm{BD}}+\sin\theta|1_{\rm{A}}\rangle_{\rm{BD}}|1_{\rm B}\rangle_{\rm{BD}},
\end{eqnarray}
where $\theta\in(0,\frac{\pi}{2})$.
Now let Alice stay in the global chart, while Bob is restricted to the  $L$  region of the de Sitter open chart. From Eqs.(\ref{w9}) and (\ref{w10}), we find that the vacuum and one-particle excitation states seen by the observer in the global chart become the entangled states in the open charts. Thus Eq.(\ref{w12}) can be written as
\begin{eqnarray}\label{w13}
\nonumber|\Psi\rangle_{\rm{AB_{L}B_{R}}}&=&\cos\theta\sqrt{1-|\gamma_{p}|^{2}}\sum_{n=0}^{\infty}\gamma_{p}^{n}|0\rangle_{\rm{BD}} |n \rangle_{\rm L} |n\rangle_{\rm R}\\
&+&\frac{\sin\theta(1-|\gamma_{p}|^{2})}{\sqrt{2}}\sum_{n=0}^{\infty}\gamma_{p}^{n}\sqrt{n+1}|1\rangle_{\rm{BD}}[|n+1 \rangle_{\rm L} |n\rangle_{\rm R}+|n\rangle_{\rm L} |(n+1)\rangle_{\rm R}].
\end{eqnarray}
For simplicity, hereafter, we omit the indices $\rm A$ and $\rm B$ in $|n_{\rm A}\rangle$ and $|n_{\rm B}\rangle$, unless there may be any confusion.

Both Alice and Bob have detectors which can detect respectively the mode $\rm A$ and mode $\rm B$ of the entangled state.
Since Bob is causally disconnected from the state in the $R$ region of the de Sitter open chart, we should trace over the degrees of freedom in the open chart $R$ and obtain
\begin{eqnarray}\label{w14}
\rho_{\rm{AB_{L}}}=(1-|\gamma_{p}|^{2})\sum_{n=0}^{\infty}|\gamma_{p}|^{2n}\rho_{\rm{AB_{L}}}^n,
\end{eqnarray}
where
\begin{eqnarray}\label{w144}
\nonumber\rho_{\rm{AB_{L}}}^n&=&\cos^{2}\theta|0 n\rangle\langle 0 n|+\sin^{2}\theta\frac{1-|\gamma_{p}|^{2}}{2}(n+1)(|1 n\rangle\langle1 n|+|1 n+1\rangle\langle1 n+1|)\\
\nonumber &+&\cos\theta\sin\theta\sqrt{\frac{1-|\gamma_{p}|^{2}}{2}}\sqrt{n+1}(\gamma_{p}|0 n+1\rangle\langle1  n|+\gamma_{p}^*|1 n\rangle\langle0 n+1|)\\
\nonumber&+&\cos\theta\sin\theta\sqrt{\frac{1-|\gamma_{p}|^{2}}{2}}\sqrt{n+1}(|0 n\rangle\langle1 n+1|+|1 n+1\rangle\langle0 n|)\\
&+&\sin^{2}\theta\frac{1-|\gamma_{p}|^{2}}{2}\sqrt{(n+1)(n+2)}(\gamma_{p}|1 n+2\rangle\langle1 n|+\gamma_{p}^{*}|1 n\rangle\langle1 n+2|),
\end{eqnarray}
where $|n m\rangle=|n\rangle_{\rm{BD}}|m\rangle_{\rm{L}}$.
Therefore, the $l_1$-norm coherence of state $\rho_{\rm{AB_{L}}}$ is
\begin{eqnarray}\label{w16}
C_{l_1}(\rho_{\rm{AB_{L}}})&=&(1-|\gamma_{p}|^{2})\sum_{n=0}^{\infty}|\gamma_{p}|^{2n}[\sin2\theta\sqrt{\frac{1-|\gamma_{p}|^{2}}{2}}(|\gamma_{p}|+1)\sqrt{n+1}\\ \nonumber
&+&\sin^{2}\theta|\gamma_{p}|(1-|\gamma_{p}|^{2})\sqrt{(n+1)(n+2)}].
\end{eqnarray}

In the bases $\{|0, n\rangle,  |0, n+1\rangle, |0, n+2\rangle, |1, n\rangle,  |1, n+1\rangle, |1, n+2\rangle\}$, we can rewrite Eq.(\ref{w144}) in the matrix,
\begin{eqnarray}\label{wq22}
\rho_{\rm{AB_{L}}}^n=\left(\!\!\begin{array}{cccccccc}
\rho_{11}^n & 0 & 0 & 0 & \rho_{15}^n & 0\\
0 & 0 & 0 & \rho_{24}^n & 0 & 0\\
0 & 0 & 0 & 0 & 0 & 0\\
0 &\rho_{42}^n & 0 & \rho_{44}^n & 0&
\rho_{46}^n\\
\rho_{51}^n & 0 & 0 & 0 & \rho_{55}^n & 0\\
0 & 0 & 0 &\rho_{64}^n & 0 & 0\\
\end{array}\!\!\right),
\end{eqnarray}
 where the matrix elements are written by
\begin{eqnarray}
\nonumber \rho_{11}^n&=&\cos^{2}\theta,\\
\nonumber \rho_{44}^n&=&\rho_{55}^n=\sin^{2}\theta\frac{(1-|\gamma_{p}|^{2})(1+n)}{2},\\
 \rho_{15}^n&=&\rho_{51}^n=\sin2\theta\frac{\sqrt{(1-|\gamma_{p}|^{2})(n+1)}}{2\sqrt{2}},\\
\nonumber \rho_{24}^n&=&(\rho_{42}^{n })^{*}=\sin2\theta\frac{\sqrt{(1-|\gamma_{p}|^{2})(n+1)} \gamma_{p}}{2\sqrt{2}},\\
\nonumber \rho_{46}^n&=&(\rho_{64}^{n})^{*}=\sin^{2}\theta\frac{(1-|\gamma_{p}|^{2})\sqrt{(n+1)(n+2)}\gamma_{p}^{*}}{2}.
\end{eqnarray}
The  matrix of Eq.(\ref{wq22}) has three nonzero eigenvalues
\begin{eqnarray}\label{ZS25}
\lambda_{1,n}=\frac{1}{4}\big\{3+n-|\gamma_{p}|^{2}-n|\gamma_{p}|^{2}+[1+|\gamma_{p}|^{2}+n(|\gamma_{p}|^{2}-1)]\cos2\theta\big\},
\end{eqnarray}
\begin{eqnarray}\label{ZS2l5}
\nonumber \lambda_{2/3,n}&=&\frac{1}{8}\bigg\{1+n-|\gamma_{p}|^{2}-n|\gamma_{p}|^{2}+\cos2\theta(1+n)(|\gamma_{p}|^{2}-1)\bigg\}\\
& \pm & \frac{\sqrt{2}}{8}\bigg\{(1+n)(|\gamma_{p}|^{2}-1)\bigg[1+n+15|\gamma_{p}|^{2}+3n|\gamma_{p}|^{2}-8|\gamma_{p}|^{4}-4n|\gamma_{p}|^{4}\\\nonumber
&+&\cos2\theta[-1+|\gamma_{p}|^{2}+8|\gamma_{p}|^{4}+n(-1-3|\gamma_{p}|^{2}+4|\gamma_{p}|^{4})]\bigg]\sin^{2}\theta\bigg\}^{\frac{1}{2}}.
\end{eqnarray}
Thus, the relative entropy of coherence  of state $\rho_{\rm{AB_{L}}}$  is
\begin{eqnarray}\label{w16}
\nonumber C_{RE}(\rho_{\rm{AB_{L}}})&=&\sum_{n=0}^{\infty}(1-|\gamma_{p}|^{2})|\gamma_{p}|^{2n}\big\{\lambda_{1,n}\log_2[(1-|\gamma_{p}|^{2})|\gamma_{p}|^{2n}\lambda_{1,n}]\\\nonumber &+&\lambda_{2,n}\log_2[(1-|\gamma_{p}|^{2})|\gamma_{p}|^{2n}\lambda_{2,n}]
+\lambda_{3,n}\log_2[(1-|\gamma_{p}|^{2})|\gamma_{p}|^{2n}\lambda_{3,n}]\\
&-&\sin^{2}\theta{(1-|\gamma_{p}|^{2})(1+n)}\log_2[\sin^{2}\theta|\gamma_{p}|^{2n}\frac{(1-|\gamma_{p}|^{2})^2(1+n)}{2}]\\ \nonumber &-&\cos^{2}\theta\log_2[\cos^{2}\theta(1-|\gamma_{p}|^{2})|\gamma_{p}|^{2n}]\big\}.
\end{eqnarray}
Obviously, the quantum coherence $C_{l_1}(\rho_{\rm{AB_{L}}})$ and $C_{RE}(\rho_{\rm{AB_{L}}})$ depends on not only the initial parameter $\theta$, but also
the curvature parameter $p$ and mass parameter $\nu$ of the de
Sitter space.

\begin{figure}
\begin{minipage}[t]{0.5\linewidth}
\centering
\includegraphics[width=3in]{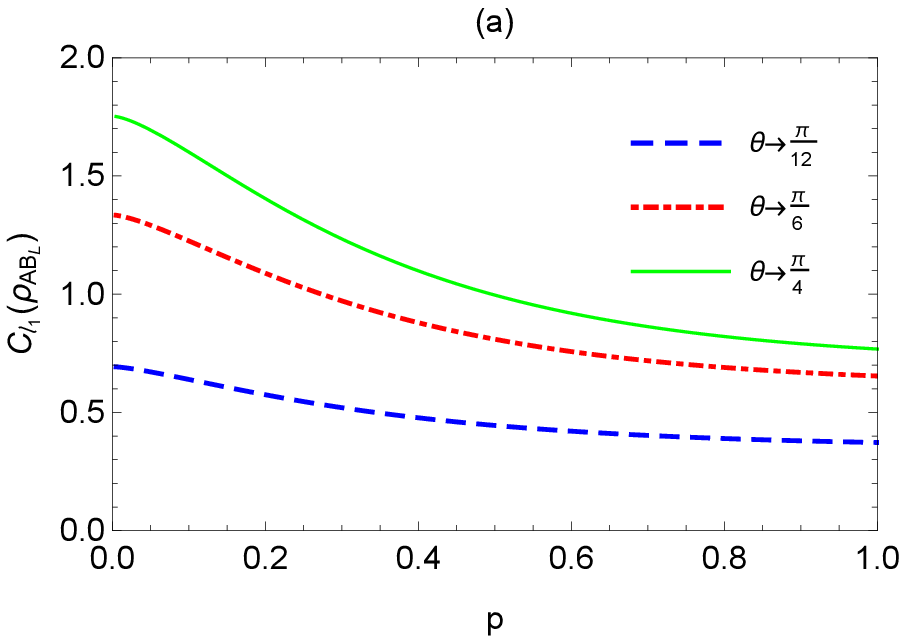}
\label{fig:side:a}
\end{minipage}%
\begin{minipage}[t]{0.5\linewidth}
\centering
\includegraphics[width=3in]{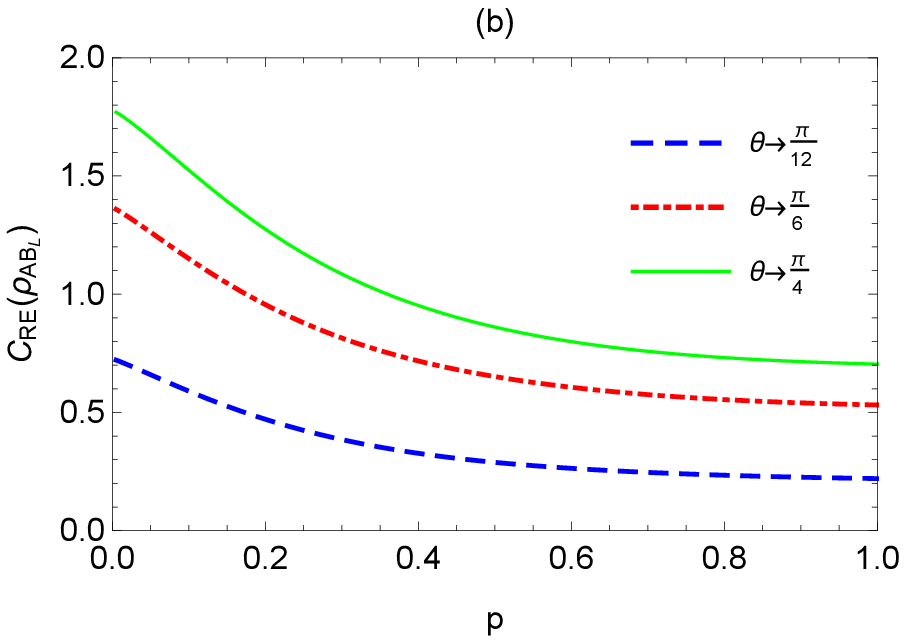}
\label{fig:side:a}
\end{minipage}%

\caption{Quantum coherence $C_{l_1}(\rho_{\rm{AB_{L}}})$ and $C_{RE}(\rho_{\rm{AB_{L}}})$ as functions of the curvature parameter $p$ for different $\theta$. The mass parameter $\nu$ is fixed as $\nu=1/2$.
 }
\label{Fig2}
\end{figure}

In Fig\ref{Fig2}, we plot quantum coherence $C_{l_1}(\rho_{\rm{AB_{L}}})$ and $C_{RE}(\rho_{\rm{AB_{L}}})$ as functions of the curvature parameter $p$ of de Sitter space for different $\theta$. It shows that both $C_{l_1}(\rho_{\rm{AB_{L}}})$ and $C_{RE}(\rho_{\rm{AB_{L}}})$  increases monotonically  with the decrease of curvature parameter $p$ in de Sitter space, meaning that space curvature enhances quantum coherence of the system of Alice and Bob. This result is opposite to the quantum correlation (entanglement and discord) which decreases monotonically under the curvature effect of de Sitter space \cite{LL42,L42,L43}. The different behaviors reflect different characteristics between quantum coherence and quantum correlation.

Note that, in Rindler spacetime, both quantum correlation and coherence decrease monotonically with the increase of acceleration \cite{L21,L22,L23,L24,L25,L26,L27,L28,L29,L30,L31,L32,L33,L34}. This means that the Unruh effect and the curvature effect of de Sitter space play similar roles to quantum correlation (destroy quantum correlation), but play different roles to quantum coherence (Unruh effect destroys, while the curvature effect of de Sitter space improves quantum coherence).
Also note that the behaviors of quantum resources (entanglement, discord and coherence) in the background of black holes are similar to that in Rindler spacetime \cite{QWE81,QWE82,L35,L36,L37,L38,L39,L40}.

Quantum coherence in a bipartite system may be distinguished as the local and
correlated \cite{LL5}. The local coherence belongs to each single subsystem, which comes from the coherent superposition between the
levels of the particular subsystem; While the correlated coherence reflects the quantum correlation between the two subsystems, which cannot be attributed to particular subsystems. The correlated coherence for a bipartite system $A$ and $B$ is defined as
\begin{eqnarray}\label{wwwcc}
C^{cc}(\rho_{AB})&=&C(\rho_{AB})-C(\rho_A)-C(\rho_{B}),
\end{eqnarray}
where $C(\rho_{AB})$ denotes the total coherence of the bipartite system, and $C(\rho_A)$ and $C(\rho_B)$ are the local coherence for subsystems $A$ and $B$, respectively. What we have studied in the above is the total quantum coherence of the system $\rm {AB_{L}}$. Note that the initial state given by Eq.(\ref{w12}) has only correlated coherence, and the local coherence of each subsystem is zero.
We wonder if this kind of distribution of coherence remains unchanged under the curvature effect of de Sitter space. So we continue to study the local coherence of each systems and their correlated coherence. By tracing over the modes $\rm{B_L}$ or $\rm A$ in the state $\rho_{\rm{AB_{L}}}$, we obtain the reduced state of each subsystem
\begin{eqnarray}\label{w25}
\rho_{\rm A}=\cos^2\theta|0\rangle\langle0|+\sin^2\theta|1\rangle\langle1|,
\end{eqnarray}
\begin{eqnarray}\label{w26}
\rho_{\rm{B_{L}}}=(1-|\gamma_{p}|^{2})\sum_{n=0}^{\infty}|\gamma_{p}|^{2n}\rho_{\rm{B_{L}}}^n,
\end{eqnarray}
where
\begin{eqnarray}\label{w26}
 \nonumber\rho_{\rm{B_{L}}}^n&=&\cos^{2}\theta+\sin^{2}\theta\frac{1-|\gamma_{p}|^{2}}{2}(n+1)]| n\rangle\langle  n|\\ \nonumber
&+&\sin^{2}\theta\frac{1-|\gamma_{p}|^{2}}{2}(n+1)|n+1\rangle\langle n+1|\\
&+&\sin^{2}\theta\frac{1-|\gamma_{p}|^{2}}{2}\sqrt{(n+1)(n+2)}(\gamma_{p}|n+2\rangle\langle n|+\gamma_{p}^{*}|n\rangle\langle n+2|).
\end{eqnarray}
 The matrix representation of $\rho_{\rm{B_{L}}}^n$ in the subspace $\{|\rm n\rangle,  |\rm n+1\rangle, |\rm n+2\rangle\}$ is given by
\begin{eqnarray}\label{w2A2}
\rho_{\rm{B_{L}}}^n=\left(\!\!\begin{array}{cccccccc}
\cos^{2}\theta+\sin^{2}\theta\frac{(1-|\gamma_{p}|^{2})(1+n)}{2}&0&\sin^{2}\theta\frac{(1-|\gamma_{p}|^{2})\sqrt{(n+1)(n+2)}\gamma_{p}^*}{2}\\
0&\sin^{2}\theta\frac{(1-|\gamma_{p}|^{2})(1+n)}{2}&0\\
\sin^{2}\theta\frac{(1-|\gamma_{p}|^{2})\sqrt{(n+1)(n+2)}\gamma_{p}}{2}&0&0
\end{array}\!\!\right),
\end{eqnarray}
which has three nonzero eigenvalues:
\begin{eqnarray}\label{w26m}
\nonumber \lambda_{4/5,n}&=&\frac{1}{16}\bigg\{6+2n-2|\gamma_{p}|^{2}-2n|\gamma_{p}|^{2}+2\cos2\theta-2n\cos2\theta+2|\gamma_{p}|^{2}\cos2\theta
\\\nonumber
&+&2n|\gamma_{p}|^{2}\cos2\theta\pm\sqrt{2}\big\{19+10n+3n^{2}+(14+20n+6n^{2})|\gamma_{p}|^{2}
-45|\gamma_{p}|^{4}\\
&-&66n|\gamma_{p}|^{4}-21n^{2}|\gamma_{p}|^{4}+24|\gamma_{p}|^{6}+36n|\gamma_{p}|^{6}
+12n^{2}|\gamma_{p}|^{6}-4[-3+6|\gamma_{p}|^{2}\\\nonumber
&-&15|\gamma_{p}|^{4}+8|\gamma_{p}|^{6}+n^{2}(-1+|\gamma_{p}|^{2})^{2}(1+4|\gamma_{p}|^{2})+2n
(-1+|\gamma_{p}|^{2})^{2}\\\nonumber
&\times&(1+6|\gamma_{p}|^{2})]\cos2\theta+[1+10|\gamma_{p}|^{2}-15|\gamma_{p}|^{4}+8|\gamma_{p}|^{6}+n^{2}(-1+|\gamma_{p}|^{2})^{2}\\
&\times&(1+4|\gamma_{p}|^{2})
+2n(-1+6|\gamma_{p}|^{2}-11|\gamma_{p}|^{4}+6|\gamma_{p}|^{6})]\cos4\theta\big\}^{\frac{1}{2}}\bigg\}\nonumber,
\end{eqnarray}
\begin{eqnarray}\label{wwwcc8}
\lambda_{6,n}=-\frac{\sin^{2}\theta}{2}(1+n)(-1+|\gamma_{p}|^{2}).
\end{eqnarray}
From these, we get the local and the correlated coherences as
\begin{eqnarray}\label{w27}
C_{l_1}(\rho_{\rm A})=C_{RE}(\rho_{\rm A})=0,
\end{eqnarray}
\begin{eqnarray}\label{w28}
C_{l_1}(\rho_{\rm{B_{L}}})=\sin^{2}\theta|(1-|\gamma_{p}|^{2})^2\sum_{n=0}^{\infty}|\gamma_{p}|^{2n+1}\sqrt{(n+1)(n+2)},
\end{eqnarray}
\begin{eqnarray}\label{w16op}
\nonumber C_{RE}(\rho_{\rm{B_{L}}})&=&\sum_{n=0}^{\infty}(1-|\gamma_{p}|^{2})|\gamma_{p}|^{2n}\bigg\{\lambda_{4,n}\log_2[(1-|\gamma_{p}|^{2})|\gamma_{p}|^{2n}\lambda_{4,n}]\\\nonumber &+&\lambda_{5,n}\log_2[(1-|\gamma_{p}|^{2})|\gamma_{p}|^{2n}\lambda_{5,n}]
+\lambda_{6,n}\log_2[(1-|\gamma_{p}|^{2})|\gamma_{p}|^{2n}\lambda_{6,n}]\\\nonumber
&-&[\cos^{2}\theta+\sin^{2}\theta\frac{(1-|\gamma_{p}|^{2})(1+n)}{2}]\log_2\big\{(1-|\gamma_{p}|^{2})|\gamma_{p}|^{2n}\\
&\times&[\cos^{2}\theta+\sin^{2}\theta\frac{(1-|\gamma_{p}|^{2})(1+n)}{2}]\big\}\\ \nonumber &-&\sin^{2}\theta\frac{(1-|\gamma_{p}|^{2})(1+n)}{2}\log_2[\sin^{2}\theta|\gamma_{p}|^{2n}\frac{(1-|\gamma_{p}|^{2})^2(1+n)}{2}]\bigg\},
\end{eqnarray}
\begin{eqnarray}\label{www28}
C_{l_1}^{cc}(\rho_{\rm A\rm{B_{L}}})&=&\sin2\theta\sqrt{\frac{(1-|\gamma_{p}|^{2})^{3}}{2}}(1+|\gamma_{p}|)\sum_{n=0}^{\infty}|\gamma_{p}|^{2n}\sqrt{n+1},
\end{eqnarray}
and
\begin{eqnarray}\label{www28U}
C_{RE}^{cc}(\rho_{\rm A\rm{B_{L}}})=C_{RE}(\rho_{\rm A\rm{B_{L}}})-C_{RE}(\rho_{\rm{B_{L}}}).
\end{eqnarray}
Obviously, the observer Bob in the $L$ region of the open chart has local quantum coherence, meaning that the curvature effect of de Sitter space can generate local coherence.
This is in sharp contrast with the Unruh effect which cannot produce local coherence [see appendix B].

\begin{figure}
\begin{minipage}[t]{0.5\linewidth}
\centering
\includegraphics[width=3in]{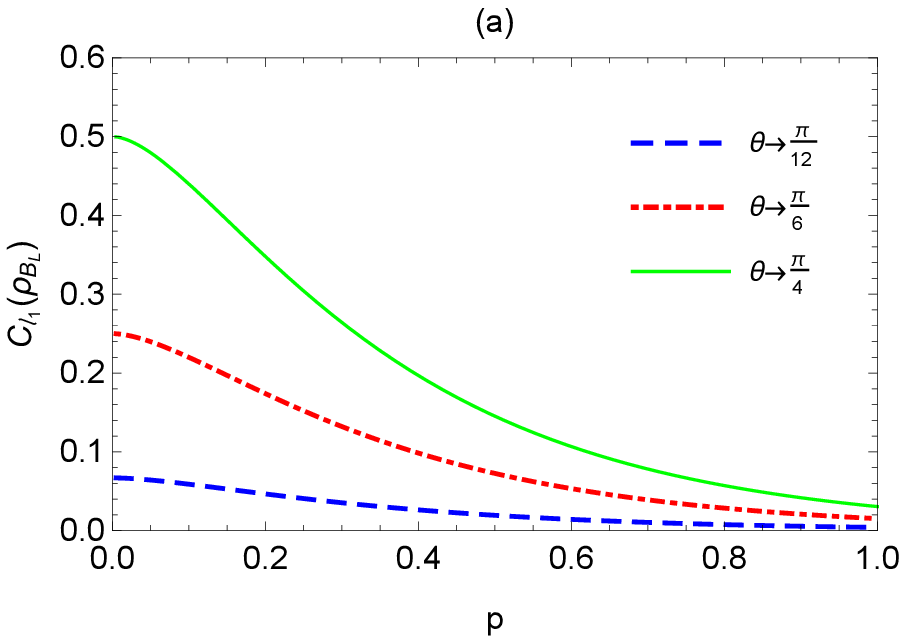}
\label{fig:side:a}
\end{minipage}%
\begin{minipage}[t]{0.5\linewidth}
\centering
\includegraphics[width=3in]{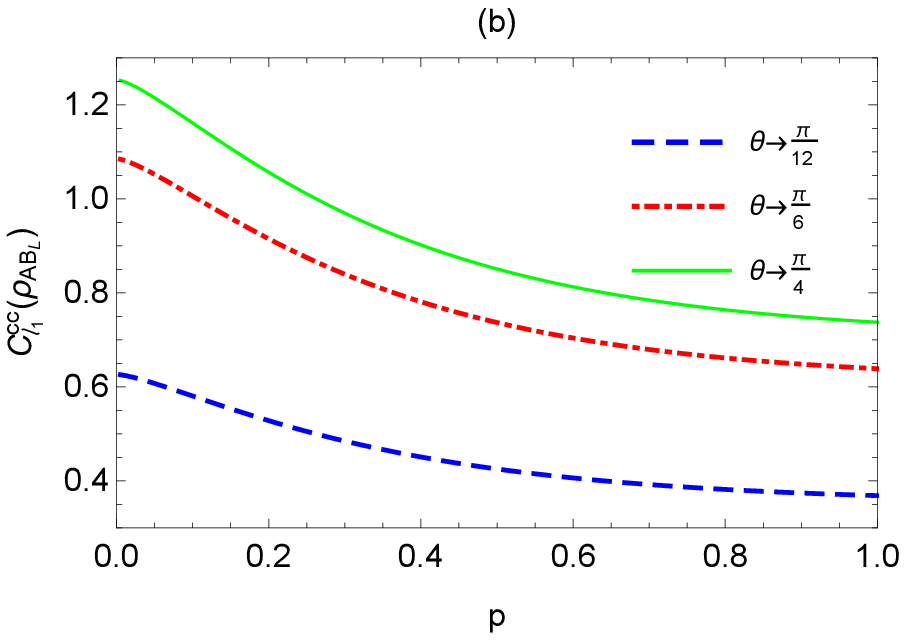}
\label{fig:side:a}
\end{minipage}%

\begin{minipage}[t]{0.5\linewidth}
\centering
\includegraphics[width=3in]{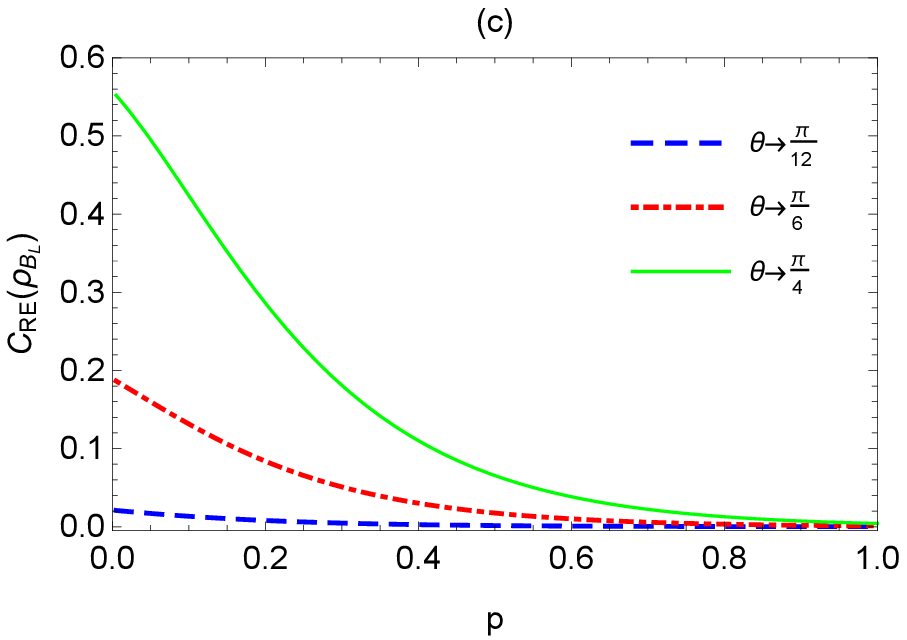}
\label{fig:side:a}
\end{minipage}%
\begin{minipage}[t]{0.5\linewidth}
\centering
\includegraphics[width=3in]{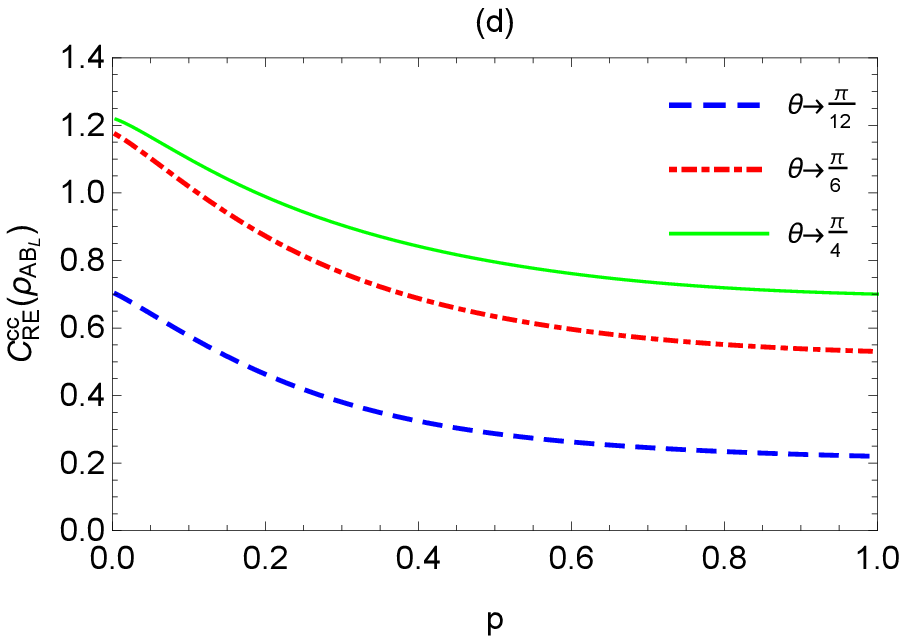}
\label{fig:side:a}
\end{minipage}%

\caption{Local quantum coherence  and correlated coherence of state $\rho_{\rm{AB_{L}}}$ as functions of the curvature parameter $p$ for different $\theta$. The mass parameter $\nu$ is fixed as $\nu=1/2$.
 }
\label{Fig3}
\end{figure}
\begin{figure}[htbp]
\centering
\includegraphics[height=1.8in,width=2.0in]{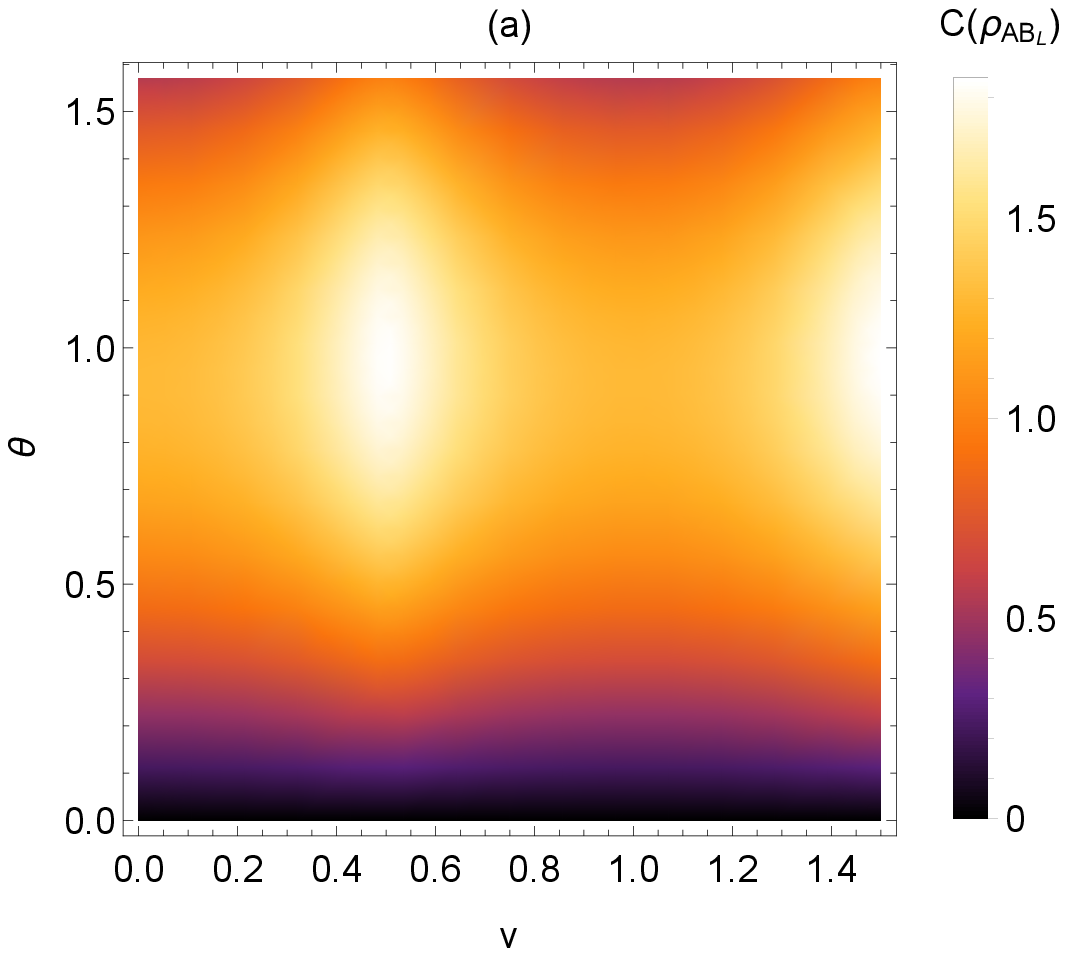}
\includegraphics[height=1.8in,width=2.0in]{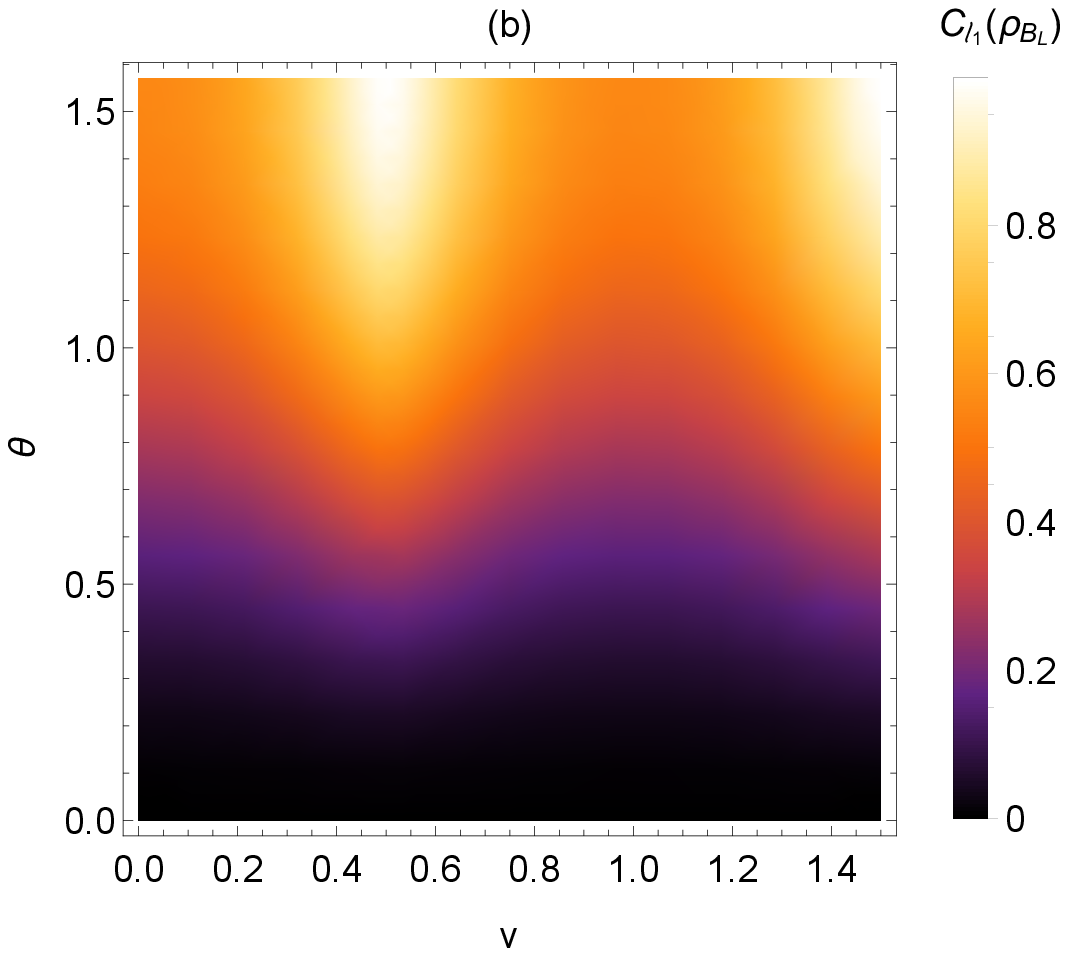}
\includegraphics[height=1.8in,width=2.0in]{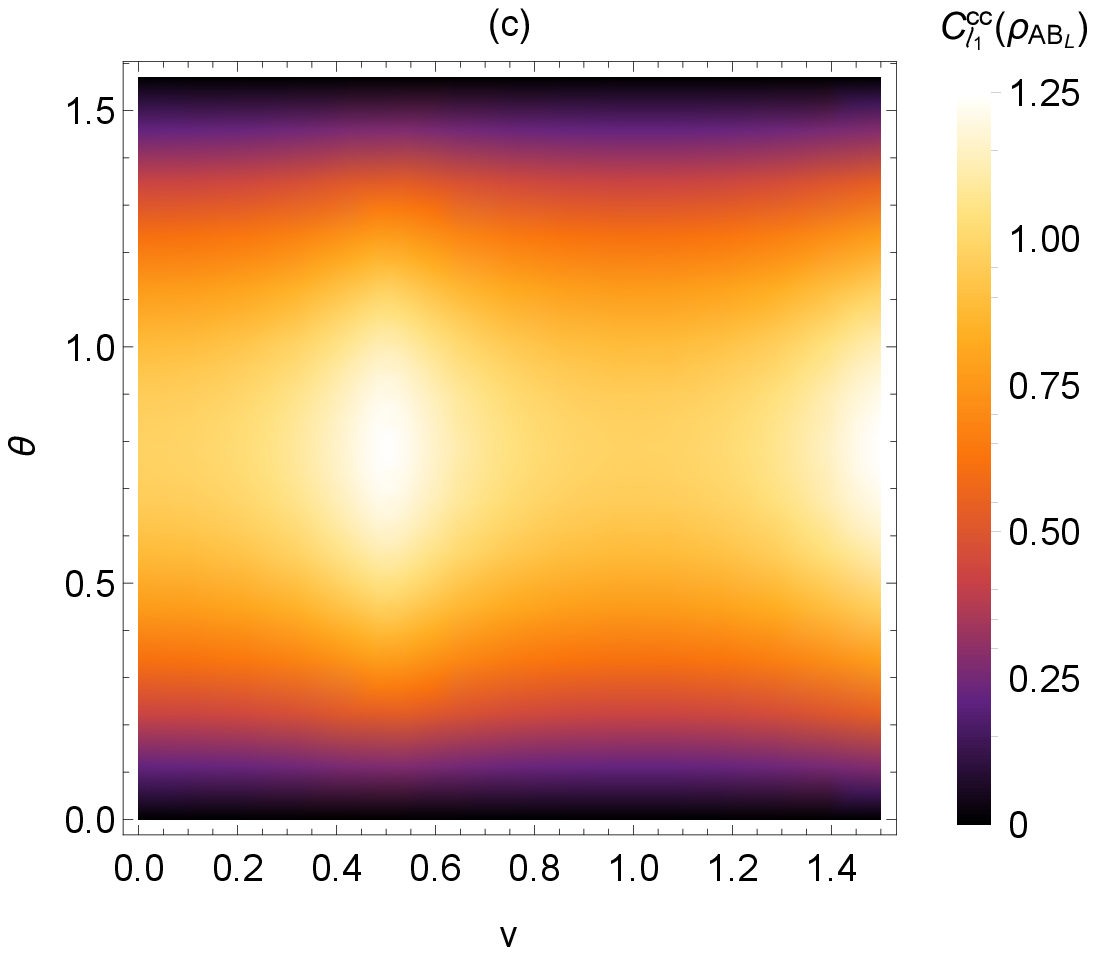}

\includegraphics[height=1.8in,width=2.0in]{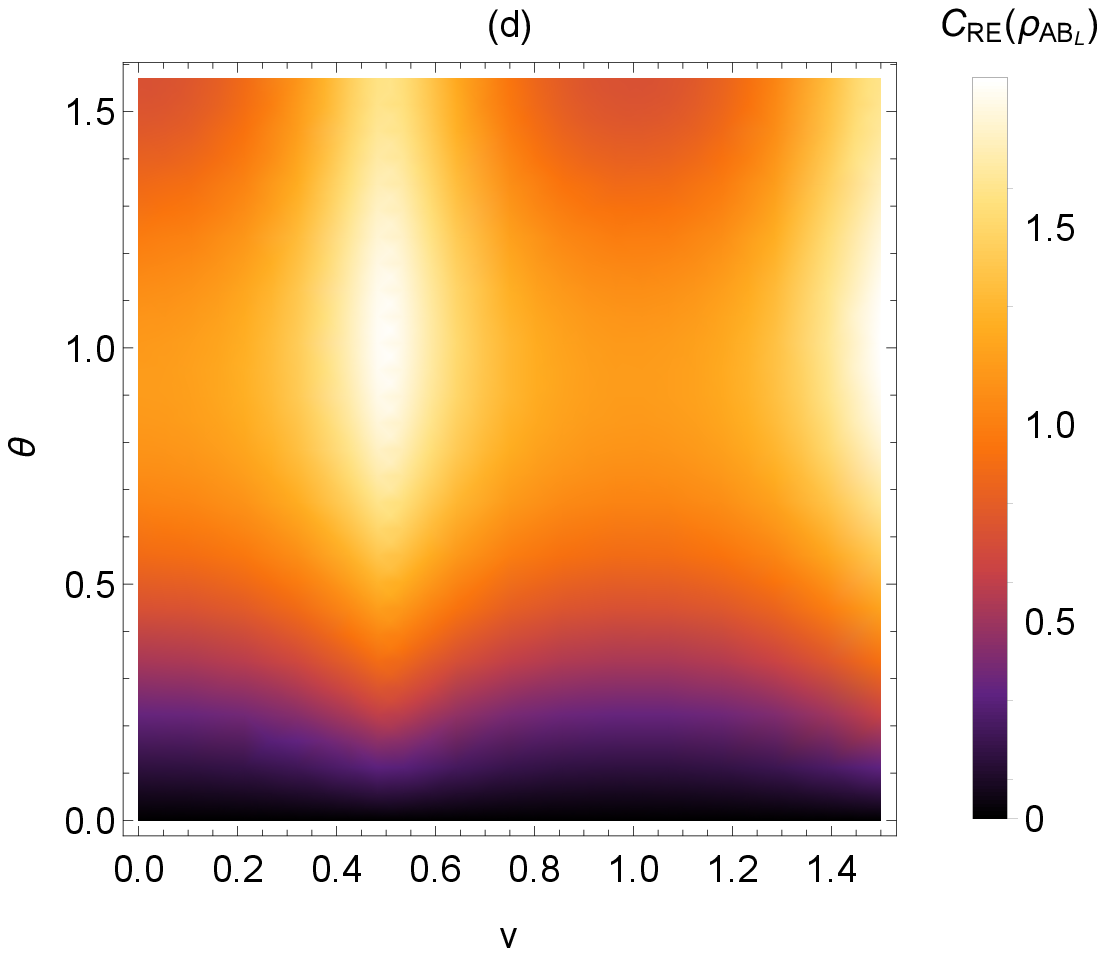}
\includegraphics[height=1.8in,width=2.0in]{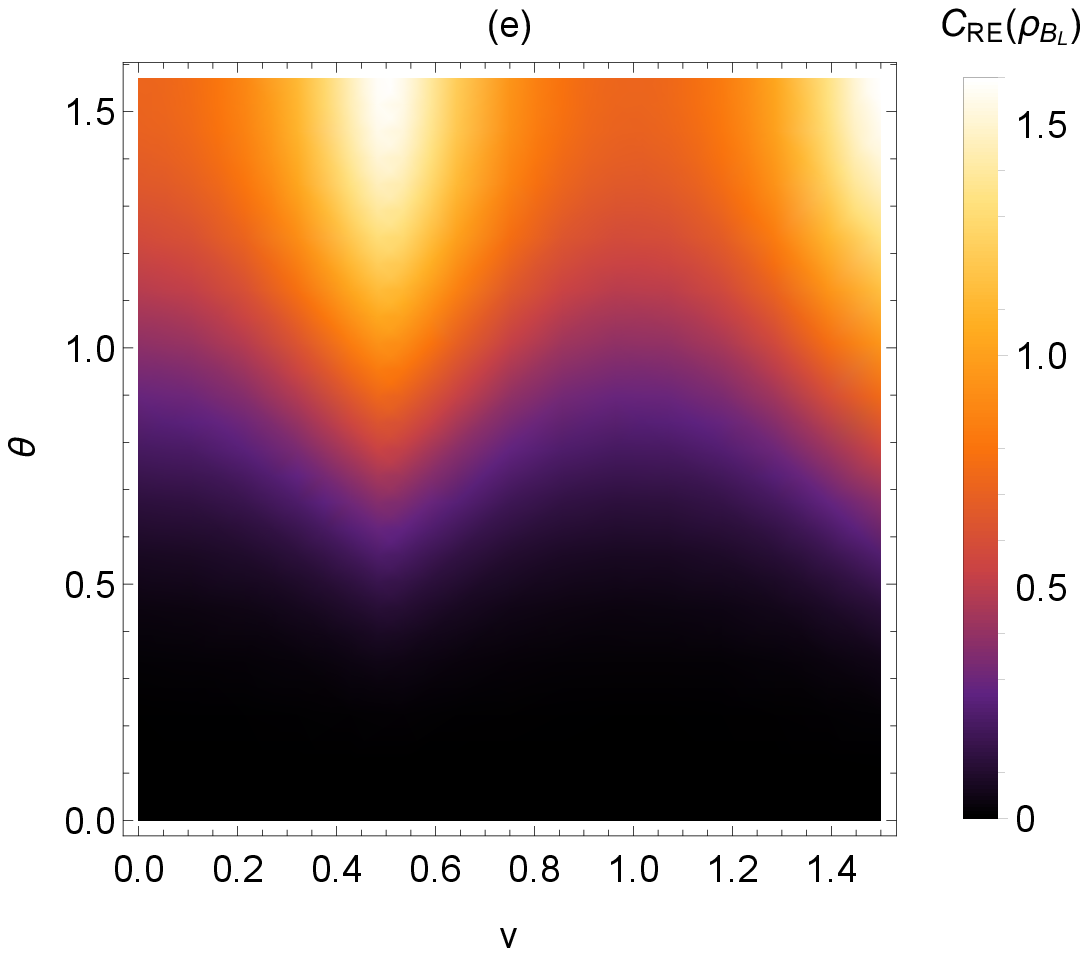}
\includegraphics[height=1.8in,width=2.0in]{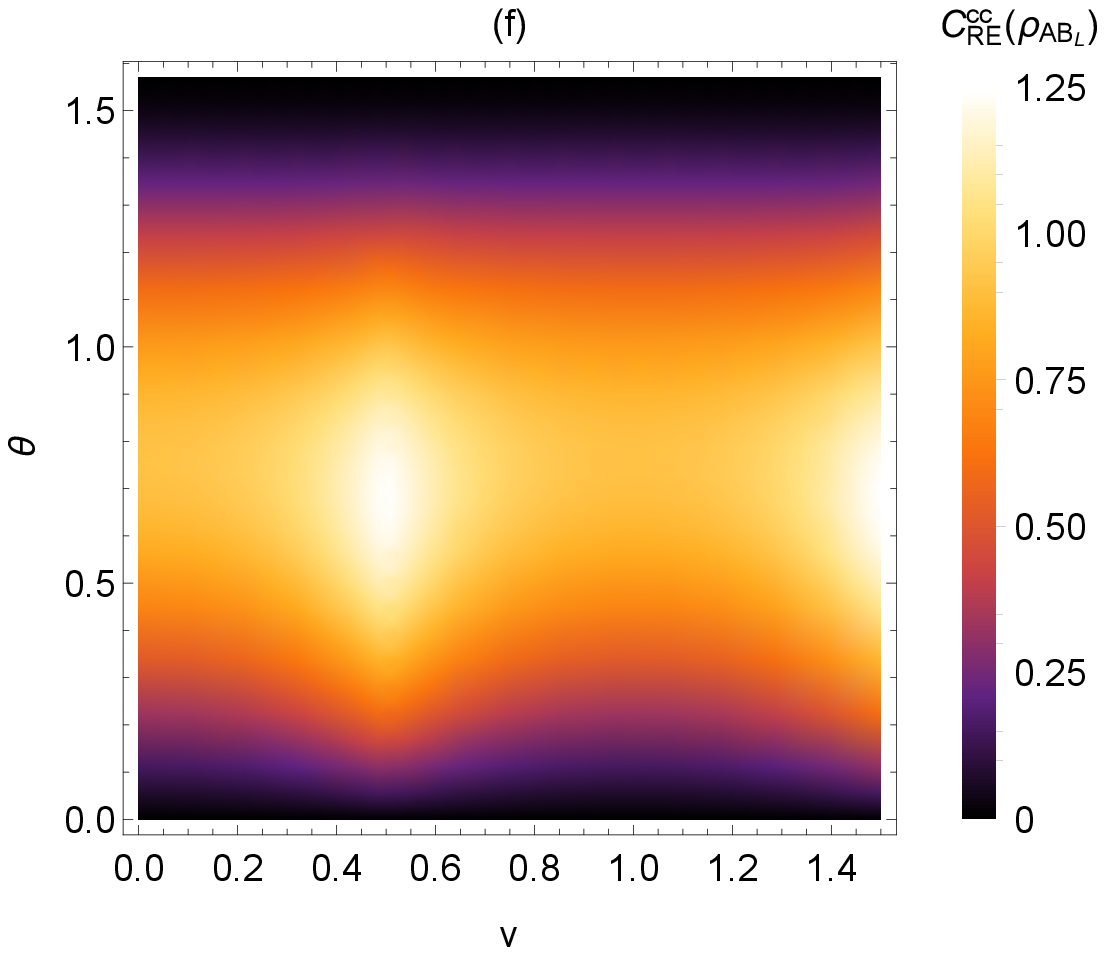}
\caption{ The total coherence, local coherence  and correlated coherence of state $\rho_{\rm{AB_{L}}}$ as functions of the mass parameter $\nu$ and the initial  parameter $\theta$.
 The curvature parameter $p$ is fixed as $p=0$.}\label{Fig4}
\end{figure}

In Fig\ref{Fig3}, we plot the local quantum coherence  and correlated coherence as functions of the curvature parameter $p$ for different $\theta$.
It is shown that both local quantum coherence  and correlated coherence of quantum state $\rho_{\rm{AB_{L}}}$ increase monotonically with the decrease of parameter $p$, meaning that curvature effect of de Sitter space can not only produce the local coherence but also enhance the correlated coherence of the underlying system.
Note that Unruh effect and Hawking effect neither produce local coherence, nor enhance the correlated coherence in Rindler and black hole spacetimes [see appendix B].

In Fig\ref{Fig4}, we plot the total coherence, local coherence and correlated coherence of state $\rho_{\rm{AB_{L}}}$
as functions of the mass parameter $\nu$ and the initial  parameter $\theta$.
It is shown that quantum coherence is most severely affected by the curvature
of de Sitter space for the cases $\nu=1/2$ (conformal invariance) and $\nu=3/2$ ( masslessness).
With the growth of the $\theta$, the total coherence $C_{l_1}(\rho_{\rm{AB_{L}}})$ and $C_{RE}(\rho_{\rm{AB_{L}}})$, and the correlated coherence $C_{l_1}^{cc}(\rho_{\rm A\rm{B_{L}}})$ and $C_{RE}^{cc}(\rho_{\rm A\rm{B_{L}}})$ increase from zero to the maximum and then decrease, while the local coherence $C_{l_1}(\rho_{\rm{B_{L}}})$ and $C_{RE}(\rho_{\rm{B_{L}}})$ increases monotonically. Note that these results are consistent with the results shown in Fig\ref{Fig2} and Fig\ref{Fig3}.

In order to facilitate the physical analysis, we rewrite the expression of
the $l_1$-norm of coherence of state $\rho_{\rm{AB_{L}}}$
\begin{eqnarray}\label{Qw16}
C_{l_1}(\rho_{\rm{AB_{L}}})=\sin2\theta\alpha(p,\nu)+\sin^2\theta\beta(p,\nu),
\end{eqnarray}
where $\alpha(p,\nu)=\sqrt{\frac{(1-|\gamma_{p}|^{2})^{3}}{2}}(1+|\gamma_{p}|)\sum_{n=0}^{\infty}|\gamma_{p}|^{2n}\sqrt{n+1}$
and $\beta(p,\nu)=(1-|\gamma_{p}|^{2})^2\sum_{n=0}^{\infty}|\gamma_{p}|^{2n+1}\sqrt{(n+1)(n+2)}$.
Naturally, we obtain
\begin{eqnarray}\label{Qw116}
C_{l_1}^{cc}(\rho_{\rm A\rm{B_{L}}})=\sin2\theta\alpha(p,\nu),\quad C_{l_1}(\rho_{\rm{B_{L}}})=\sin^2\theta\beta(p,\nu) \quad \theta\in(0,\frac{\pi}{2}).
\end{eqnarray}
From Eqs.(\ref{Qw16}) and (\ref{Qw116}), we see that the correlated coherence $C_{l_1}^{cc}(\rho_{\rm A\rm{B_{L}}})$, local coherence $C_{l_1}(\rho_{\rm{B_{L}}})$ and total coherence $C_{l_1}(\rho_{\rm{AB_{L}}})$ include three variables, $\theta$, $p$ and $\nu$. Now, we study the behavior of these coherences when only one of the variables changes and the other two variables remain unchanged. Firstly, when $\theta$ changes in the area $(0, \pi/2)$, the local coherence $C_{l_1}(\rho_{\rm{B_{L}}})$ changes monotonically, and the correlated coherence $C_{l_1}^{cc}(\rho_{\rm A\rm{B_{L}}})$ and total coherence $C_{l_1}(\rho_{\rm{AB_{L}}})$ each have a peak. Secondly, both $\alpha(p,\nu)$ and $\beta(p,\nu)$ are the increasing functions of $|\gamma_p|$, and thus the decreasing functions of $p$. Therefore, the three coherences are all the decreasing functions of $p$, meaning that the space curvature leads to the increase of the coherences. Finally, both $\nu=1/2$ and $\nu=3/2$ are the maximum points of $\alpha(p,\nu)$ and $\beta(p,\nu)$. Thus, the three coherences appear peaks for the cases $\nu=1/2$ (conformal invariance) and $\nu=3/2$ ( masslessness). This analysis is in agreement with Fig\ref{Fig4} and clearly shows the influences of curvature effect and mass parameter on quantum coherence.

The mass parameter $\nu$  of the scalar
field in de Sitter space depends on the two scales: the mass $m$ and the Hubble radius $H^{-1}$. For $\nu=1/2$ ($m^2/H^2=2$), the system is conformally invariant.
We first consider the meaning of the Bunch-Davies vacuum for the case $\nu=1/2$.
The $\alpha$-vacuum for the case $\nu=1/2$ is $\alpha$-independent and is equivalent to Bunch-Davies vacuum. This is because that conformally flat space is indistinguishable from Minkowski space for the case $\nu=1/2$ \cite{LL44}. Then,
the Legendre function for the case $\nu=1/2$ reduces to an elementary function as \begin{eqnarray}\label{S3ui0}
P^{ip}_{0}(\cosh t)=\frac{e^{-ip\eta}}{ip\Gamma(-ip)},
\end{eqnarray}
where $\eta$ is related to $t$ by the relation, $\sinh \eta=\frac{-1}{\sinh t}$.
Since we focus on the  $L$ region, a natural choice of the positive frequency function is ${\chi}_{p}=\frac{e^{-ip\eta}}{\sqrt{2p}}$, which means that the
Bunch-Davies vacuum state is seen as a thermal state \cite{L45}.
In other words, for the case $\nu=1/2$, the reduced density matrix of Eq.(\ref{qws119}) reduces to a thermal state with temperature $T=H/(2\pi)$ and the generated entanglement entropy is maximum \cite{LL42,LLL44}.  Similarly, for the case $\nu=1/2$, it is beneficial to the curvature effect to generate larger quantum coherence. In addition, quantum coherence in de Sitter space depends on the oscillatory behavior that comes from the factor $\cos2\pi\nu$ in $|\gamma_p|$.
For fixed curvature parameter $p$, $|\gamma_p|$ is maximum for the case $\nu=1/2$.
Therefore,  quantum coherence is most severely affected by the curvature effect of de Sitter space for the cases of conformal invariance ($\nu=1/2$).

From the above analysis, we see that quantum coherence in de Sitter space has completely different behavior compared with that in the Rindler spacetime or in the black hole spacetime. In the following, we want from the perspective of Bogoliubov transformation to clarify how the different behaviors have been produced. As the Bogoliubov transformations in the black hole spacetime and in the Rindler spacetime are similar, we thus take the Rindler spacetime as the example. Unfortunately, we can present the discussion only based on the $l_1$-norm of coherence, because of its simplicity.
The Minkowski vacuum and excitation states observed by an inertial observer relate to the states of $I$ and $II$ regions in Rindler spacetime as follows \cite{L21,L28}
\begin{eqnarray}\label{w17}
|0\rangle_{\rm M}=\frac{1}{\cosh (r)}\sum_{n=0}^\infty\tanh^n(r)|n\rangle_{\rm I}|n\rangle_{\rm{II}},
\end{eqnarray}
\begin{eqnarray}\label{w18}
|1\rangle_{\rm M}=\frac{1}{\cosh^2 (r)}\sum_{n=0}^\infty\tanh^n(r)\sqrt{n+1}|n+1\rangle_{\rm I}|n\rangle_{\rm {II}},
\end{eqnarray}
with $r$ the acceleration parameter. We find that this transformation is asymmetric with respect to the  $I$ and $II$ regions. However, the transformation given by Eqs.(\ref{w9}) and (\ref{w10}) is symmetric with respect to the  $R$ and $L$ regions. It is this symmetry and asymmetry of transformations that lead to different behaviors of quantum coherence. Specifically, Eqs.(\ref{w9}) and (\ref{w10}) lead to the evolution of the density matrices from Bunch-Davies basis to the basis of the open chart $L$ region as
\begin{eqnarray}\label{w19}
\rm{Tr_R(|0\rangle_{BD}\langle0|)}=(1-|\gamma_{p}|^{2})\sum_{n=0}^{\infty}|\gamma_{p}|^{2n}| n\rangle_{\rm L}\langle  n|,
\end{eqnarray}
\begin{eqnarray}\label{w20}
\nonumber\rm{Tr_R(|0\rangle_{BD}\langle1|)}&=&(1-|\gamma_{p}|^{2})^{\frac{3}{2}}\sum_{n=0}^{\infty}|\gamma_{p}|^{2n}
\sqrt{\frac{n+1}{2}}
\\
&&(\gamma_{p}| n+1\rangle_{\rm L}\langle  n| +| n\rangle_{\rm L}\langle n+1|),
\end{eqnarray}
\begin{eqnarray}\label{w21}
\nonumber\rm{Tr_R(|1\rangle_{BD}\langle1|)}&=&\frac{(1-|\gamma_{p}|^{2})^2}{2}\sum_{n=0}^{\infty}|\gamma_{p}|^{2n}[(n+1)(| n\rangle_{\rm L}\langle n|+| n+1\rangle_{\rm L}\langle n+1|)\\
&+&\sqrt{(n+1)(n+2)}(\gamma_{p}| n+2\rangle_{\rm L}\langle n|+\gamma_{p}^{*}| n\rangle_{\rm L}\langle n+2|)].
\end{eqnarray}
Similarly, Eqs.(\ref{w17}) and (\ref{w18}) lead to the evolution of the density matrices from Minkowski basis to the basis of $I$ region of Rindler spacetime given by
\begin{eqnarray}\label{w22}
\rm{Tr_{II}(|0\rangle_{M}\langle0|)}=\frac{1}{\cosh^2 r}\sum_{n=0}^\infty\tanh^{2n}r|n\rangle_{\rm I}\langle n|,
\end{eqnarray}
\begin{eqnarray}\label{w23}
\rm{Tr_{II}(|0\rangle_{M}\langle1|)}=\frac{1}{\cosh^3 r}\sum_{n=0}^\infty\tanh^{2n}r\sqrt{n+1}|n\rangle_{\rm I}\langle n+1|,
\end{eqnarray}
\begin{eqnarray}\label{w24}
\rm{Tr_{II}(|1\rangle_{M}\langle1|)}=\frac{1}{\cosh^4 r}\sum_{n=0}^\infty\tanh^{2n}r(n+1)|n+1\rangle_{\rm I}\langle n+1|.
\end{eqnarray}
By comparing Eqs.(\ref{w22})-(\ref{w24}) with Eqs.(\ref{w19})-(\ref{w21}), we find the difference between the evolutions of density matrices: The curvature effect of de Sitter space can make diagonal elements of the density matrix in the Bunch-Davies basis, $(|1\rangle_{\rm {BD}}\langle1|)$, to produce non-diagonal elements in the basis of the open chart $L$ region [see Eq.(\ref{w21})]. Further, the produced non-diagonal elements have different positions with the non-diagonal elements produced by $(|0\rangle_{\rm{BD}}\langle1|)$ and its conjugation [see Eq.(\ref{w20})]. Obviously, the diagonal element $(|1\rangle_{\rm{BD}}\langle1|)$ always has positive contribution to the $l_1$-norm of quantum coherence via curvature effect of de Sitter space. This positive contribution can clearly produce the local coherence $C_{l_1}(\rho_{\rm{B_{L}}})$. Actually it is also the reason for enhancing the correlated coherence $C_{l_1}^{cc}(\rho_{\rm A\rm{B_{L}}})$ as shown in Fig\ref{Fig3}(b). Therefore the total coherence increases under the curvature effect of de Sitter space. In contrast, the acceleration effect cannot make diagonal elements of the density matrix in the Minkowski basis to produce non-diagonal elements in the Rindler-$I$ spacetime [see Eqs.(\ref{w22}) and (\ref{w24})], and the coherence in the Rindler-$I$ spacetime comes completely from the evolution of the non-diagonal elements $(|0\rangle_{\rm M}\langle1|)$ and its conjugation. Thus Unruh effect cannot produce the local $l_1$-norm of coherence. As the correlated coherence under the Unruh effect degrades (see appendix B),
the total coherence also degrades.

\section{ Conclusions  \label{GSCDGE}}
Based on the two measures of $l_1$-norm of coherence and relative entropy of coherence, the effect of space curvature of de Sitter space on quantum coherence of an entangled state
between two free modes of a massive scalar field has been investigated.
We assume that the one mode (named by mode $\rm A$) of the scalar field is observed by Alice who is in the global chart, and another mode (named by mode $\rm{B_L}$) is observed by Bob who is in the $L$ region of the open chart of de Sitter space. We have shown that the curvature effect of de Sitter space on the one hand can produce local coherence of Bob, and on the other hand can enhance the correlated coherence between Alice and Bob, so that the total coherence of the bipartite system of Alice and Bob increases monotonically with the increase of curvature. This makes sharp contrast with the quantum correlation (quantum entanglement and discord) which decreases monotonically with the increase of curvature of de Sitter space \cite{L42,L43}. For the first time, we have found that quantum coherence and quantum correlation show completely opposite behavior in the framework of relativity.

We have also compared the influences of the curvature effect of de Sitter space with Unruh effect or Hawking effect on quantum coherence. We have found that Unruh effect and Hawking effect cannot produce local coherence and at the same time destroy the correlated coherence, so that the total coherence of the bipartite system of Alice and Bob decreases monotonically with the increase of acceleration in Rindler spacetime \cite{L28,L29} or with the strengthening of Hawking radiation in the black hole spacetime \cite{L40}. Therefore, quantum coherence exhibits the opposite trend in de Sitter space comparing to Rindler spacetime or black hole spacetime.
Because de Sitter space is the unique maximally symmetric curved space, the transformation given by Eqs.(\ref{w9}) and (\ref{w10}) is symmetric with respect to the  $R$ and $L$ regions of de Sitter space. However, the transformation given by Eq.(\ref{w18}) is asymmetric with respect to the  $I$ and $II$ regions in Rindler spacetime. Symmetry of de Sitter space leads to curvature effect that can increase quantum coherence, while asymmetry of Rindler spacetime  leads to Unruh effect that can decrease quantum coherence.

The dynamics of the underlying quantum coherence in de Sitter space encodes much information about the expansion of Universe, which may help us to understand the behaviors of expansion of Universe. We expect our results can play roles in the prediction of cosmological observations, and in the simulating of the Universe expansion in superconducting circuit and in an ultracold quantum fluid of light \cite{L45m, L45m1}.

\begin{acknowledgments}
This work is supported by the National Natural
Science Foundation of China (Grant Nos. 12205133, 1217050862, 11275064, 11975064 and 12075050 ), LJKQZ20222315 and 2021BSL013.	
\end{acknowledgments}

\appendix
\onecolumngrid

\section{The mode functions of de Sitter space }
We expand a free scalar field in de Sitter space
\begin{eqnarray}\label{S2pf}
\hat{\phi}(x)=\sum_\Lambda(\hat{a}_{\Lambda}u_{\Lambda}(x)+\hat{a}_{\Lambda}^{\dag}{u_{\Lambda}(x)}^*),
\end{eqnarray}
where \{$u_{\Lambda}(x)$\} and their complex conjugate form a complete set of mode functions
denoted by $\Lambda$ that satisfy the field equation
\begin{eqnarray}\label{S3jk0}
[g^{\mu\nu}\nabla_{\mu}\nabla_{\nu}-m^{2}]u_{\Lambda}(x)=0,
\end{eqnarray}
and normalized with respect to the Klein-Gordon inner product \cite{L45}.
We can determines the vacuum state $\hat{a}_{\Lambda}|0\rangle=0$ for any $\Lambda$ associated with a specified set of mode functions.

One can write down the field equation via the coordinates in either of the  $R$ or $L$  regions to find the Euclidean vacuum mode functions on the open chart of de Sitter space.
Since the $R$ and $L$ regions are completely symmetric, we obatin
\begin{eqnarray}\label{S30tv}
[\frac{1}{a^{3}(t)}\frac{\partial}{\partial t}a^{3}(t)\frac{\partial}{\partial t}-\frac{H^{-2}}{a^{2}(t)} {\rm\bf L^2}+\frac{9}{4}-\nu^{2}]u_{\sigma p\ell m}(t,r,\Omega)=0,
\end{eqnarray}
where $(t,r)=(t_{R},r_{R})$ or $(t_{L},r_{L})$, $a(t)=H^{-1}\sinh t$, $\nu=\sqrt{\frac{9}{4}-\frac{m^{2}}{H^{2}}}$ and ${\rm\bf L^2}=\frac{1}{\sinh^{2}r}\frac{\partial}{\partial r}(\sinh^{2}r\frac{\partial}{\partial r})+\frac{1}{\sinh^{2}r}{\rm\bf L^2}_{\Omega}$.
Here, ${\rm\bf L^2}_{\Omega}$ denotes the usual Laplacian operator on the unit two-sphere.
Then  the eigenvalue equation for the operator $-{\rm\bf L^2}$ on a unit three-dimensional hyperboloid can be given by
\begin{eqnarray}\label{S30df}
-{\rm\bf L^2}Y_{plm}(r,\Omega)=(1+p^{2})Y_{plm}(r,\Omega),
\end{eqnarray}
where the eigenfunction $Y_{plm}$ that is regular at $r=0$ can be written as
\begin{eqnarray}\label{S3k0}
Y_{plm}(r,\Omega)&=&f_{pl}(r)Y_{lm}(\Omega),
\notag\\f_{pl}(r)&:=&\frac{\Gamma(ip+l+1)}{\Gamma(ip+1)}\frac{p}{\sqrt{\sinh r}}P^{-l-\frac{1}{2}}_{ip-\frac{1}{2}}(\cosh r)\notag\\&=&(-1)^{l}\sqrt{\frac{2}{\pi}}\frac{\Gamma(-ip+1)}{\Gamma(-ip+l+1)}\sinh^{l} r\frac{d^{l}}{d(\cosh r)^{l}}(\frac{\sin pr}{\sinh r}),
\end{eqnarray}
where $P^{\nu}_{\mu}(z)$ is the associated Legendre function of the first kind, $\Gamma(z)$ is the gamma function and $Y_{lm}(\Omega)$ is the normalized spherical harmonic function on the unit two-sphere \cite{L45}.
For real positive values of $p$,
the eigenfunctions $Y_{plm}$  form an orthonormal complete
set for squareintegrable functions on the unit three-hyperboloid, and are normalized as
\begin{eqnarray}\label{S3e0}
\int^{\infty}_{0}dr \sinh^{2}r\int d\Omega Y_{plm}(r,\Omega){Y^*_{p'l'm'}(r,\Omega)}\notag\\
=\delta(p-p')\delta_{ll'}\delta mm'.
\end{eqnarray}
Now the harmonic expansion of the mode functions can be given by
\begin{eqnarray}\label{S305g}
u_{\sigma p\ell m}(t,r,\Omega)=\frac{H}{\sinh t}\chi_{\sigma plm}(t)Y_{plm}(r,\Omega).
\end{eqnarray}

\section{Quantum coherence in  Rindler spacetime }
In this appendix, we present the influence of Unruh effect on quantum coherence. Assume that Alice and Bob initially share a state $|\Phi\rangle_{\rm {AB}}$ [given by Eq.(\ref{w12})] of bosonic fields in inertial frame. Then Alice stays in the inertial frame, while Bob  moves with a uniform acceleration. According to the transformation of Eqs.(\ref{w17})-(\ref{w18}), we get
\begin{eqnarray}\label{A1}
|\Psi\rangle_{\rm{AB_IB_{II}}}=\frac{\cos\theta}{\cosh r}\sum_{n=0}^\infty\tanh^nr|0nn\rangle
+\frac{\sin\theta}{\cosh^2r}\sum_{m=0}^\infty\tanh^mr\sqrt{m+1}|1(m+1) m\rangle,
\end{eqnarray}
where we abbreviate $|i_{\rm A}j_{\rm{B_I}}k_{\rm{B_{II}}}\rangle$ as $|ijk\rangle$.
Tracing over the inaccessible mode ${\rm B_{II}}$ in region $II$, we obtain the density matrix
\begin{eqnarray}\label{A2}
\rho_{\rm{AB_I}}=\frac{1}{\cosh^2 r}\sum_{n=0}^\infty(\tanh^{2n}r)\rho_{\rm{AB_I}}^n,
\end{eqnarray}
where
\begin{eqnarray}\label{A22}
\rho_{\rm{AB_I}}^n &&=\cos^2\theta|0n\rangle\langle0n|+\frac{\sin^2\theta(n+1)}{\cosh^2r}|1n+1\rangle\langle1n+1|
\\
\nonumber &&+\frac{\sin2\theta\sqrt{n+1}}{2\cosh r}|0n\rangle\langle1n+1|+\frac{\sin2\theta\sqrt{n+1}}{2\cosh r}|1n+1\rangle\langle0n|.
\end{eqnarray}
In the subspace of $\{|0, \rm n\rangle,  |0, \rm  n+1\rangle, |1,\rm n\rangle, |1,\rm n+1 \rangle\}$, $\rho_{\rm{AB_I}}^n$ has its matrix
\begin{eqnarray}\label{A88}
 \rho_{\rm{AB_I}}^n= \left(\!\!\begin{array}{cccccccc}
\cos^2\theta&0&0&\frac{\sin2\theta\sqrt{n+1}}{2\cosh r}\\
0&0&0&0\\
0&0&0&0\\
\frac{\sin2\theta\sqrt{n+1}}{2\cosh r}&0&0&\frac{\sin^2\theta(n+1)}{\cosh^2r}
\end{array}\!\!\right),
\end{eqnarray}
with nonzero eigenvalues:
\begin{eqnarray}\label{OA22}
\frac{\tanh^{2n}}{\cosh^2 r}\bigg[\cos^2\theta+\frac{\sin^2\theta(1+n^2)}{\cosh^2 r}  \bigg].
\end{eqnarray}

\begin{figure}
\begin{minipage}[t]{0.5\linewidth}
\centering
\includegraphics[width=3in]{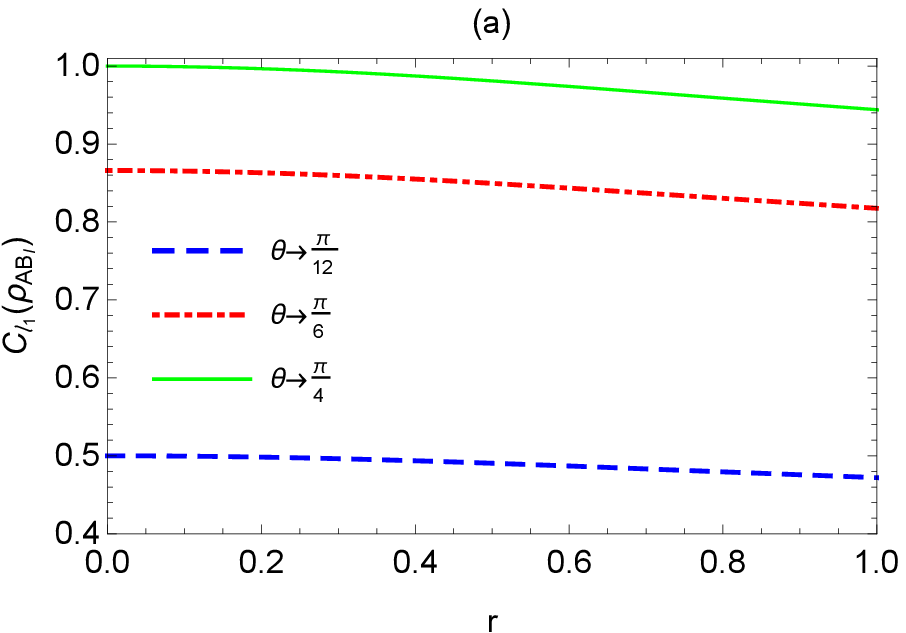}
\label{fig:side:a}
\end{minipage}%
\begin{minipage}[t]{0.5\linewidth}
\centering
\includegraphics[width=3in]{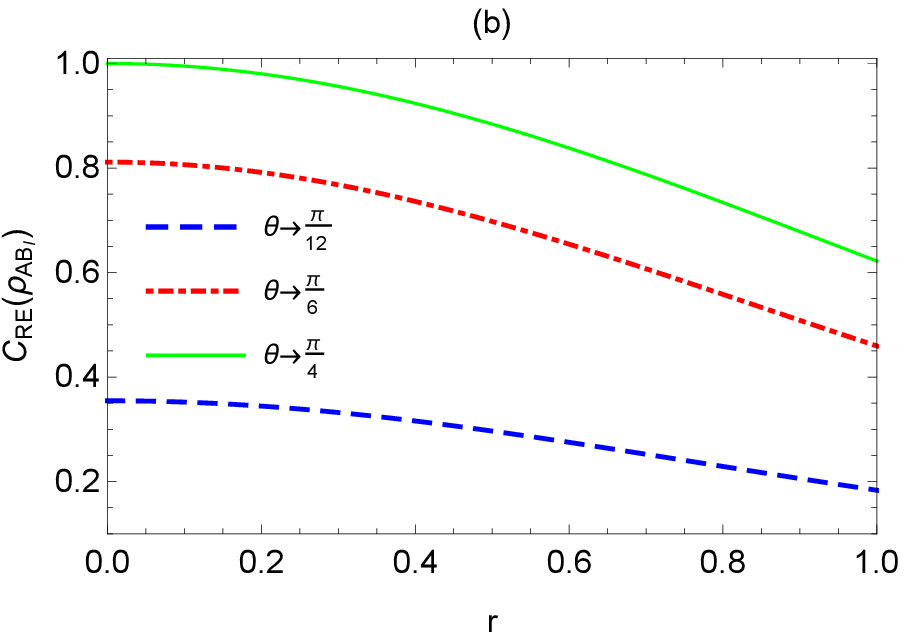}
\label{fig:side:a}
\end{minipage}%

\caption{Quantum coherence $C_{l_1}(\rho_{\rm{AB_I}})$ and $C_{RE}(\rho_{\rm{AB_I}})$ as a function of acceleration parameter $r$ for different $\theta$.
 }
\label{Fig5}
\end{figure}

According to Eqs.(\ref{w15}) and (\ref{w155}), the coherence of this state in  Rindler spacetime can be expressed as
\begin{eqnarray}\label{A3}
 C_{l_1}(\rho_{\rm{AB_I}})=\frac{\sin2\theta}{\cosh^{3}r} \sum_{n=0}^\infty\sqrt{n+1}\tanh^{2n}r,
\end{eqnarray}
\begin{eqnarray}\label{Aq3}
\nonumber C_{RE}(\rho_{\rm{AB_I}})&=&\sum_{n=0}^\infty\bigg\{\frac{\tanh^{2n}}{\cosh^2 r}\bigg[\cos^2\theta+\frac{\sin^2\theta(1+n^2)}{\cosh^2 r}  \bigg]\log_2\bigg[\frac{\tanh^{2n}}{\cosh^2 r}\bigg(\cos^2\theta\\ \nonumber
&+&\frac{\sin^2\theta(1+n^2)}{\cosh^2 r}\bigg)\bigg]-\frac{\cos^2\theta\tanh^{2n}}{\cosh^2 r}\log_2\bigg[\frac{\cos^2\theta\tanh^{2n}}{\cosh^2 r}\bigg]\\
&-&\frac{\sin^2\theta\tanh^{2n}(1+n^2)}{\cosh^4 r}\log_2\bigg[\frac{\sin^2\theta\tanh^{2n}(1+n^2)}{\cosh^4 r}\bigg]\bigg\}.
\end{eqnarray}
In Fig.\ref{Fig5}, we plot the coherence $C_{l_1}(\rho_{\rm{AB_I}})$ and $C_{RE}(\rho_{\rm{AB_I}})$ as a function of the acceleration parameter $r$. It is shown that the coherence reduces
monotonously with the increase of the acceleration parameter $r$. This means that the  Unruh effect reduces  the total coherence of the bipartite system of Alice and Bob.

By tracing over the mode $\rm A$ or $\rm {B_I}$ from $\rho_{\rm{AB_I}}$, we further obtain
\begin{eqnarray}\label{A4}
\rho_{\rm A}=\cos^2\theta|0\rangle\langle0|+\sin^2\theta|1\rangle\langle1|,
\end{eqnarray}
\begin{eqnarray}\label{GHZ2}
\rho_{\rm{B_I}}=\sum_{n=0}^\infty\tanh^{2n}r\left[\frac{\cos^2\theta}{\cosh^2 r}|n\rangle\langle n|+\frac{\sin^2\theta}{\cosh^4 r}(n+1)|n+1\rangle\langle n+1|\right].
\end{eqnarray}
Obviously, the local quantum  coherences for both Alice and Bob are equal to zero. Therefore, the correlated coherence is equal to the total coherence, $C^{cc}_{l_1}(\rho_{\rm{AB_I}})=C_{l_1}(\rho_{\rm{AB_I}})$ and $C^{cc}_{RE}(\rho_{\rm{AB_I}})=C_{RE}(\rho_{\rm{AB_I}})$, which also reduces
monotonously with the increase of the acceleration parameter $r$.



\begin{thebibliography}{99}
\bibitem{L1}
A. J. Leggett, Prog. Theor. Phys. Suppl. {\bf69}, 80 (1980).

\bibitem{L2}
M. A. Nielsen and I. L. Chuang, Quantum Computation and
Quantum Information, 10th ed. (Cambridge University Press,
Cambridge, 2010).

\bibitem{L3}
A. Streltsov, G. Adesso, and M. B. Plenio, Colloquium: Quantum
coherence as a resource, Rev. Mod. Phys. {\bf89}, 041003
(2017).

\bibitem{L4}
R. Horodecki, P. Horodecki, M. Horodecki, and K. Horodecki,
Quantum entanglement, Rev. Mod. Phys. {\bf81}, 865 (2009).

\bibitem{L5}
A. Einstein, B. Podolsky, and N. Rosen,
Phys. Rev. 47, {\bf777} (1935).

\bibitem{LL5}
K. Chuan Tan, H. Kwon, C. Y. Park, and H. Jeong, Phys. Rev. A {\bf94}, 022329 (2016).

\bibitem{LLL5}
Y. Guo, S. Goswami, Phys. Rev. A {\bf95}, 062340 (2017).

\bibitem{L6}
B. Schumacher, M. D. Westmoreland, Phys. Rev. Lett. {\bf80}, 5695 (1998).

\bibitem{L7}
S. E. Barnes, R. Ballou, B. Barbara, J. Strelen,
 Phys. Rev. Lett. {\bf79}, 289 (1997).


\bibitem{L9}
U. K. Sharma, I. Chakrabarty, M. K. Shukla,
 Phys. Rev. A {\bf96}, 052319 (2017).

\bibitem{L10}
Y. Peng, Y. Jiang, H. Fan,
 Phys. Rev. A {\bf93}, 032326 (2016).

\bibitem{L11}
F. G. S. L. Brand\~{a}o, M. Horodecki, N. H. Y. Ng, J. Oppenheim, and S. Wehner,
Proc. Natl. Acad. Sci. U. S. A. {\bf112}, 3275 (2015).

\bibitem{L12}
M. Horodecki and J. Oppenheim,
Nat. Commun. {\bf4}, 2059 (2013).

\bibitem{L13}
P. \'{C}wikli\'{n}ski, M. Studzi\'{n}ski, M. Horodecki, and J. Oppenheim,
Phys. Rev. Lett. {\bf115}, 210403 (2015).

\bibitem{L14}
S. F. Huelga and M. B. Plenio,
 Nat. Phys. {\bf10}, 621 (2014).

\bibitem{L15}
S. F. Huelga and M. B. Plenio, Contemp. Phys. {\bf54}, 181 (2013).


\bibitem{L16}
M. G\"{a}rttner, P. Hauke, and A. M. Rey,
Phys. Rev. Lett. {\bf 120}, 040402 (2018).

\bibitem{L17}
T. Baumgratz, M. Cramer, and M. B. Plenio,
Phys. Rev. Lett. {\bf113}, 140401 (2014).

\bibitem{L18}
W. G. Unruh, Phys. Rev. D {\bf14}, 870 (1976).

\bibitem{L19}
S. W. Hawking, Nature (London) {\bf248}, 30 (1974).

\bibitem{L20}
J. L. Ball, I. Fuentes-Schuller, F. P. Schuller, Phys. Lett. A
{\bf359}, 550 (2006).

\bibitem{L20L}
Y. Nambu and Y. Ohsumi, Phys. Rev. D
{\bf84}, 044028 (2011).


\bibitem{L20LL}
S. Kanno, J. Murugan, J. P. Shock, and J. Soda,  J. High Energy
Phys. {\bf07}, 072  (2014).


\bibitem{L21}
I. Fuentes-Schuller, and R. B. Mann, Phys. Rev. Lett. {\bf 95},120404 (2005).

\bibitem{L22}
P. M. Alsing, I. Fuentes-Schuller, R. B. Mann, and T. E.
Tessier, Phys. Rev. A {\bf74}, 032326 (2006).

\bibitem{L23}
G. Adesso, I. Fuentes-Schuller, and M. Ericsson, Phys. Rev. A {\bf76}, 062112 (2007).

\bibitem{L24}
D. E. Bruschi, J. Louko, E. Mart\'{\i}n-Mart\'{\i}nez, A. Dragan, and I. Fuentes,
Phys. Rev. A {\bf82}, 042332 (2010).

\bibitem{L25}
E. Mart\'{\i}n-Mart\'{\i}nez, L. J. Garay, and J. Le\'{o}n, Phys. Rev. D
{\bf82}, 064006 (2010).

\bibitem{L26}
D. E. Bruschi, A. Dragan, I. Fuentes, and J. Louko, Phys. Rev. D {\bf86}, 025026 (2012).


\bibitem{L27}
M. R. Hwang, D. Park, and E. Jung, Phys. Rev. A {\bf83}, 012111 (2011).

\bibitem{L28}
S. Harikrishnan, S. Jambulingam, P. P. Rohde, C. Radhakrishnan,  Phys. Rev. A {\bf105}, 052403 (2022).

\bibitem{L29}
S. M. Wu and H. S. Zeng, H. M. Cao,  Class. Quantum Grav. 38,  185007 (2021).

\bibitem{L30}
E. Mart\'{\i}n-Mart\'{\i}nez, I. Fuentes,  Phys. Rev. A {\bf83}, 052306 (2011).

\bibitem{L31}
J. Chang, Y. Kwon, Phys. Rev. A {\bf85}, 032302 (2012).


\bibitem{L32}
W. C. Qiang, G. H. Sun, Q. Dong, and S. H. Dong,  Phys. Rev. A {\bf98}, 022320 (2018).

\bibitem{L33}
A. J. Torres-Arenasa, Q. Dong, G. H. Sun, W. C. Qiang, S. H. Dong, Phys. Lett. B {\bf789}, 93 (2019).

\bibitem{L34}
J. Wang, J. Deng, and J. Jing, Phys. Rev. A {\bf81}, 052120 (2010).

\bibitem{QWE81}
S. M. Wu and H. S. Zeng, Eur. Phys. J. C {\bf82}, 4  (2022).

\bibitem{QWE82}
S. M. Wu, Y. T. Cai, W. J. Peng, H. S. Zeng, Eur. Phys. J. C {\bf82}, 412 (2022).

\bibitem{L35}
Q. Pan, J. Jing, Phys. Rev. D {\bf78}, 065015 (2008).

\bibitem{L36}
B. N. Esfahani, M. Shamirzaie, and M. Soltani, Phys. Rev. D {\bf84}, 025024 (2011).

\bibitem{L37}
J. He, S. Xu, L. Ye,  Phys. Lett. B {\bf756}, 278 (2016).

\bibitem{L38}
S. Xu, X. K. Song, J. D. Shi, and L. Ye, Phys. Rev. D {\bf89}, 065022 (2014).

\bibitem{L39}
J. Wang, J. Jing, and H. Fan, Phys. Rev. D {\bf90}, 025032 (2014).

\bibitem{L40}
J. Shi, J. Chen, J. He, T. Wu, and L. Ye,   Quantum Inf. Process. {\bf18}, 300 (2019).




\bibitem{LL42}
S. Kanno, J. P. Shock, and J. Soda,  J. Cosmol. Astropart. Phys. {\bf03},
015 (2015).


\bibitem{L42}
S. Kanno, J. P. Shock and J. Soda, Phys. Rev. D  {\bf94}, 125014 (2016).

\bibitem{L43}
J. Wang, C. Wen, S. Chen, J. Jing, Phys. Lett. B {\bf800}, 135109 (2020).

\bibitem{Lqw43}
C. Wen, J. Wang,  J. Jing, Eur. Phys. J. C {\bf80}, 78 (2020).

\bibitem{L44}
J. Maldacena and G. L. Pimentel,  J. High Energy Phys. {\bf02}, 038  (2013).


\bibitem{LL44}
S. Kanno, J. Murugan, J. P. Shock, and J. Soda,  J. High Energy
Phys. {\bf07}, 072  (2014).


\bibitem{LLL44}
A. Albrecht, S. Kanno, and M. Sasaki, Phys. Rev. D
{\bf97}, 083520 (2018).


\bibitem{L45}
M. Sasaki, T. Tanaka, and K. Yamamoto,
Phys. Rev. D {\bf51}, 2979 (1995).

\bibitem{Scully}
M.O. Scully, M.S. Zubairy. Quantum optics, (Cambridge University Press, Cambridge, 1997)

\bibitem{Gunter}
G. Mahler, and V.A. Weberru{\ss}. Quantum Networks, (Springer-Verlag Berlin Heidelberg, 1995)

\bibitem{Xu2020}
H. Xu, F. Xu, T. Theurer, D. Egloff, Z. W. Liu, N. Yu, M.B. Plenio, and L. Zhang, Phys. Rev. Lett. {\bf 125}, 060404 (2020).

\bibitem{Wang2017}
Y. T. Wang, J. S. Tang, Z. Y. Wei, S. Yu, Z. J. Ke, X. Y. Xu, C. F. Li, G. C. Guo, Phys. Rev. Lett. {\bf118}, 020403 (2017).

\bibitem{L45m}
Z. Tian, J. Jing, A. Dragan, Phys. Rev. D {\bf95}, 125003 (2017).


\bibitem{L45m1}
J. Steinhauer, $et$ $al$, Nature Communications {\bf13}, 2890 (2022).


\end{thebibliography}
\end{document}